\documentclass[twocolumn,amsmath,amssymb,10pt]{revtex4-2}
\pdfoutput=1
\usepackage{amsfonts}
\usepackage{amsmath}
\usepackage{amsfonts}

\usepackage{amssymb}
\usepackage{bm}
\usepackage{dcolumn}
\usepackage{epsfig}
\usepackage{graphicx}
\usepackage{graphics}
\usepackage[utf8]{inputenc}
\usepackage[braket]{qcircuit}
\usepackage[export]{adjustbox}
\usepackage{graphicx}
\usepackage{multirow}
\usepackage{caption}
\usepackage{subcaption}
\usepackage{color}

\DeclareMathOperator*{\argmin}{arg\,min}

\begin{document}

\title{Quantum variational solving of the Wheeler-DeWitt equation}

\author{Grzegorz Czelusta}%
\author{Jakub Mielczarek}%
\email{jakub.mielczarek@uj.edu.pl}
\affiliation{Institute of Theoretical Physics, Jagiellonian University, 
{\L}ojasiewicza 11, 30-348 Cracow, Poland}

\begin{abstract}
One of the central difficulties in the quantization of the gravitational 
interactions is that they are described by a set of constraints. The 
standard strategy for dealing with the problem is the Dirac quantization 
procedure, which leads to the Wheeler-DeWitt equation. However, 
solutions to the equation are known only for specific symmetry-reduced 
systems, including models of quantum cosmology. Novel methods,
which enable solving the equation for complex gravitational configurations
are, therefore, worth seeking.   

Here, we propose and investigate a new method of solving the 
Wheeler-DeWitt equation, which employs a variational quantum 
computing approach, possible to implement on quantum computers. 
For this purpose, the gravitational system is regularized, by performing
spherical compactification of the phase space.  This makes the 
system's Hilbert space finite-dimensional and allows to use $SU(2)$ 
variables, which are easy to handle in quantum computing. The 
validity of the method is examined in the case of the flat de Sitter 
universe. For the purpose of testing the method, both an emulator 
of a quantum computer and the IBM superconducting quantum 
computer have been used. Advantages and limitations of the approach 
are discussed. 
\end{abstract}

\maketitle
\section{Introduction}

The Hamiltonian of General Relativity (GR) is a sum of constraints. The constraints 
are usually grouped into two sets: a single scalar (Hamiltonian) constraint $S$ 
and a three-component vector constraint $V$, so that the gravitational Hamiltonian is:
\begin{equation}
H[N,\vec{N}] = S[N]+V[\vec{N}]=\int_{\Sigma}d^3x N C + \int_{\Sigma}d^3x \vec{N} \cdot  \vec{C},
\end{equation}  
where $\Sigma$ is a spatial hypersurface. The $N$ and $\vec{N}$ are the lapse 
function and the shift vector respectively, which are integrated with the smeared 
constraints $C$ and $\vec{C}$. By imposing the constraint on the kinematical 
phase space $\Gamma_{\text{kin}}$ the physical phase space $\Gamma_{\text{phys}}$ 
is obtained. However, due to the complicated form of the constraints (specifically 
the smeared scalar constraint $C$), extraction of the physical phase space is in general 
a difficult task. 

This difficulty propagates onto the quantum case where the initial kinematical Hilbert 
space $\mathcal{H}_{\text{kin}}$ is a subject of imposing the quantum constraints 
$\hat{C}$ and $\hat{\vec{C}}$ in order to extract the physical states 
$|\Psi_{\text{phys}}\rangle$, belonging to the physical Hilbert space 
$\mathcal{H}_{\text{phys}} \subseteq \mathcal{H}_{\text{kin}}$ (the equality of 
sets $\mathcal{H}_{\text{phys}} = \mathcal{H}_{\text{kin}}$ corresponds to the 
trivial case of vanishing constraints). There are various strategies for approaching 
the problem.  

Before we proceed to reviewing the most common of them let us restrict our 
considerations to the case of a single quantum constraint $\hat{C}$ - the 
quantum Hamiltonian constraint (scalar constraint). This assumption is to simplify 
our considerations and make them more transparent. However, extension to 
the case of multiple constraints is, in principle, straightforward. Namely, 
a given method of solving the single constraint has to be applied successively. 
However, additional technical difficulties may appear due to differences in the 
functional form of the constraints. Furthermore, the case with the single constraint 
$\hat{C}$ corresponds to the case of homogenenous minisuperspace models, 
which are relevant in (quantum) cosmology. 

By extracting states which solve the quantum Hamiltonian constraint  $\hat{C}$, 
a subspace $\mathcal{H}_C$ of the kinematical Hilbert space  $\mathcal{H}_{\text{kin}}$ 
can be found. In general, the subspace $\mathcal{H}_C$, is further restricted by 
solving the quantum vector constraint $\hat{\vec{C}}$, leading to  $\mathcal{H}_{\text{phys}}$, 
such that $\mathcal{H}_C \subseteq \mathcal{H}_{\text{phys}}$. However, in the 
special case of vanishing vector constraint (which is satisfied for certain minisuperspace 
models), we have $\mathcal{H}_C= \mathcal{H}_{\text{phys}}$. Therefore, in general, 
the following sequence of weak inclusions holds:  $\mathcal{H}_{\text{phys}} \subseteq 
\mathcal{H}_C \subseteq \mathcal{H}_{\text{kin}}$, which is satisfied independently 
on whether the vector constraint is solved before or after the Hamiltonian constraint. 
 
Perhaps the most common approach to determine $\mathcal{H}_C$ is provided by the 
Dirac method of quantizing constrained systems. Here, taking the $\hat{C}$, which is a 
self-adjoint operator, one is looking for states which are annihilated by the operator, i.e. 
\begin{equation}
\hat{C}| \Psi \rangle \approx 0,   
\label{WHD}
\end{equation}
where ``$\approx$" denotes weak equality, i.e. satisfied for the states 
$| \Psi \rangle \in \mathcal{H}_C$. Eq. (\ref{WHD}) is the famous Wheeler-DeWitt 
(WDW) equation.  Solutions to the equation, which are belonging to the kernel of the 
operator $\hat{C}$, span the Hilbert space $\mathcal{H}_C = \ker{\hat{C}}$. 
The difficulty of the method lies in finding solution to the WDW equation. 
The solutions are known e.g. for certain quantum cosmological models \cite{HH,Vilenkin}.

Extraction of the physical states can alternatively be performed employing 
the \emph{group averaging} \cite{Ashtekar:1995zh} approach, which utilizes the 
projection operator $\hat{P}$.  The $\hat{P}$ is a non-unitary, but self-adjoint 
($\hat{P}^{\dagger}=\hat{P}$) and idempotent ($\hat{P}^2=\hat{P}$) operator, 
which for the case of a constraint $\hat{C}$ with a zero eigenvalue takes the 
following form:
\begin{equation}
\hat{P}=\lim_{T\rightarrow \infty} \frac{1}{2T} \int_{-T}^{T}d\tau e^{i \tau \hat{C}}. 
\end{equation}
The expression performs Dirac delta-like action on the kinematical states, projecting 
them onto the physical subspace. 

Another widely explored method of finding the physical states is provided 
by the \emph{reduced phase space} method \cite{Thiemann:2004wk}, in which one 
looks for solution of the constraints already at the classical level. For gravity,  
this is perhaps not possible do in general. However, utility of the approach has 
been shown for certain minisuperspace models (see e.g. \cite{Mielczarek:2011mx}).  While 
the $\Gamma_{\text{phys}}$ is extracted, the algebra of observables is a subject of 
quantization, leading to the physical Hilbert space $\mathcal{H}_{\text{phys}}$.  

The method we are going to study here is based on the observation made in 
Ref. \cite{Mielczarek:2021xik}. Namely, while a Hamiltonian constraint $C\approx 0$ 
is considered, the configurations satisfying the constraint can be found by 
identifying ground states of a new Hamiltonian $C^2$. A possibility of extracting 
$\Gamma_{\text{phys}}$ for a prototype classical constraint $C$ with the use 
of adiabatic quantum computing has been discussed. 

Here, we generalize the method to the quantum case and investigate its 
implementation on a universal quantum computer. The approach, utilizes 
Variational Quantum Eigensolver (VQE) \cite{Peruzzo}, which is a hybrid 
quantum algorithm. The algorithm has been widely discussed in the literature, 
in particular in the context of quantum chemistry \cite{Kandala,Cao}. 
While our VQE-based method is introduced in a general fashion, which 
does not depend on the particular form of $\hat{C}$, over the article we 
will mostly refer to the concrete case of $\hat{C}$, corresponding to a 
quantum cosmological model. The VQE will be implemented on both 
simulator of a quantum computer (employing Penny Lane \cite{PennyLane} 
and Qiskit \cite{Qiskit} tools) and on actual superconducting quantum 
computer provided by IBM \cite{IBM}.

Applying quantum computing methods unavoidably requires dealing 
with the finite systems - having finite dimensional Hilbert spaces. 
Because standard canonical quantization of gravitational system does not 
lead to finite dimensional Hilbert space representation, a procedure of 
cutting-off dimension of the Hilbert space has to be applied. 
For this purpose, we apply the recently introduced Non-linear Field 
Space Theory (NFST) \cite{Mielczarek:2016rax}, which provides a systematic 
procedure of compactifying phase spaces of the standard affine phase 
spaces. The compactification leads to finite volume of the phase space, 
and in consequence finite dimension of the Hilbert space. In case of the 
spherical compactification, of a $\mathbb{R}^2$ phase space, the control 
parameter of the cut-off is the total spin $S$, associated with the volume 
of the spherical phase space. In the large spin limit ($S\rightarrow \infty$), 
the standard case with an infinite dimensional Hilbert space is recovered. 
Depending on quantum computational resources, the value of $S$ can 
be fixed such that the corresponding Hilbert space can be represented 
with available number of logical qubits.     

Additional advantage of the method introduced in the article is that finding 
physical states employing variational method gives as explicitly operator 
(i.e. ansatz with determined parameters) which can be used to generate the 
physical states on a quantum computer. The states can be used for further 
simulations on a quantum processor. For example, transitions amplitudes 
between the states can be evaluated. On the other hand, when the physical 
states are found using analytical methods or classical numerics, the difficulty 
of constructing operator preparing a given state remains.

The organization of the article is the following. In Sec. \ref{Compact}, the 
method of regularizing the Hamiltonian constraint, employing compactification
of the phase space is introduced. The procedure is applied to the case of 
de Sitter cosmology.  Then, in Sec. \ref{VQE}, general considerations 
concerning the VQE applied to solving the Hamiltonian constraint are made.  
The qubit representation of Hamiltonian constraint introduced in Sec. \ref{Compact} 
is discussed in Sec. \ref{ExpectationValues}. In Sec. \ref{FixedSpin} the problem 
of determining fixed spin subspace of the physical Hilbert space is addressed. 
A quantum method of evaluating gradients in the VQE procedure is presented 
in Sec. \ref{Gradient}. Examples of applying the procedure for the case of 
spin $s=1$ is given in Sec. \ref{S1} and in Sec. \ref{S2} for $s=2$. The 
computational complexity considerations of the method are made in Sec. 
\ref{Complexity}. The results are summarized in Sec. \ref{Summary}.

\section{Compact phase space regularization of de Sitter model}
\label{Compact}

The initial step towards quantum variational solving of the Wheeler-DeWitt 
equation is making system's Hilbert space finite. Actually, there are theoretical 
arguments for gravitational Hilbert space being locally finite \cite{Bao:2017rnv}.
Some of the approaches quantum gravity, aim to implement this property 
while performing quantization of gravitational degrees of freedom \cite{Rovelli:2015fwa}. 
Here, we will follow a general procedure of making gravitational Hilbert space finite, 
which bases on compactification of the phase space. The approach is considered 
here as a particular, convenient way of performing regularization of quantum 
system. However, it may also play a role in formulating quantum theory of 
gravitational interaction. However, this second possibility is not explored here, 
and the method is used purely for technical reason.  

\subsection{Compact phase spaces} 

Let us recall that for a system with $m$ classical degrees of freedom, 
dimension of the phase space $\Gamma$ is dim$\Gamma=2m$.
Having the symplectic form $\omega$, defined at the phase space 
(which is a symplectic manifold), the volume of the space is $\mathcal{V} 
= \int_{\Gamma} \omega$. Following the Heisenberg uncertainty principle 
one can now estimate the number of linearly independent vectors in the 
corresponding Hilbert space as:
\begin{equation}
\dim \mathcal{H} \sim \frac{\mathcal{V}}{(2\pi \hslash)^m}. 
\label{dimH}
\end{equation} 
It should be noted that because the Heisenberg uncertainty may differ from 
the standard form while quantum gravitational degrees of freedom are consider,
the formula (\ref{dimH}) may be a subject of additional modifications. This should, 
however, not affect the general observation that dimension of the Hilbert space 
is monotonically dependent on the volume of the system's phase space. In 
consequence, finiteness of both the Hilbert space and the phase space are 
are equivalent:
\begin{equation}
\dim \mathcal{H} < \infty \Leftrightarrow \mathcal{V} < \infty. 
\end{equation}
This allows us to conclude that by performing compactification of the phase 
space, the resulting quantum system will be characterized by a finite Hilbert 
space. The observation has been recently pushed forward in Non-linear Field 
Space Theory (NFST) \cite{Mielczarek:2016rax}, with the ambition of to introduce 
compact phase space generalizations not only of mechanical systems but 
also field theories. The procedure has been so far most extensively studied 
on the case of a scalar field.     

For our purpose let us focus our attention on the case with finite number of 
classical degrees of freedom. This is in particular the situation of the so-called 
minisuperspace gravitational systems. Having the $m$ classical degrees of 
freedom the standard symplectic form (in the Darboux basis) can be written as:
\begin{equation}
\omega = \sum_{i=1}^m \omega_i = \sum_{i=1}^m dp_i \wedge dq_i,  
\end{equation} 
which is defined on the $\Gamma = \mathbb{R}^{2m}$ phase space. There 
are various ways of performing compactification of the phase space.

\subsection{Spherical phase space} 

A simple and convenient approach is to replace every $\mathbb{R}^2$ subspace 
(corresponding to a given conjugated pair $(q_i,p_i)$) with a 2-sphere, 
$\mathbb{S}^2$. So that, the total phase space, for the system having $m$ 
classical degrees of freedom, becomes $\Gamma = \mathbb{S}^{2m}$.  
This replacement is possible because the $\mathbb{S}^2$ sphere is 
a symplectic manifold, and product of symplectic manifolds is also
symplectic. 

There are two main advantages of such a choice. First, 
is the fact that $\mathbb{S}^2$ is a phase space of angular momentum (spin)
which results in easy to handle and well understood representation on
both classical and quantum level. Second, a single new parameter $S$,
associated with the volume of the phase space $\mathcal{V} = 
\int_{\mathbb{S}^2} \omega_{\mathbb{S}^2} = 4\pi S$, provides a natural control parameter 
of the regularization. The flat (affine) limit is recovered by taking the $S\rightarrow \infty$ 
limit. On the other hand, selecting a given value of $S$ precisely determines 
dimension of the Hilbert space. This is because quantization of the 
spherical phase space leads to condition $S=\hslash s$, where $s=\frac{n}{2}$, 
and $n \in \mathbb{N}$. For a given quantum number $s$, the associated 
Hilbert space $\mathcal{H}_s$ has dimension $\dim \mathcal{H}_s = 2s +1$, 
which leads to the a concrete realization of the relation (\ref{dimH}):
\begin{equation}
\dim \mathcal{H}_s = 2s +1 = \frac{2S}{\hslash}+1=\frac{\mathcal{V}}{2\pi \hslash}+1. 
\end{equation}
 
\begin{figure}[ht!]
	\centering
	\includegraphics[scale=0.25]{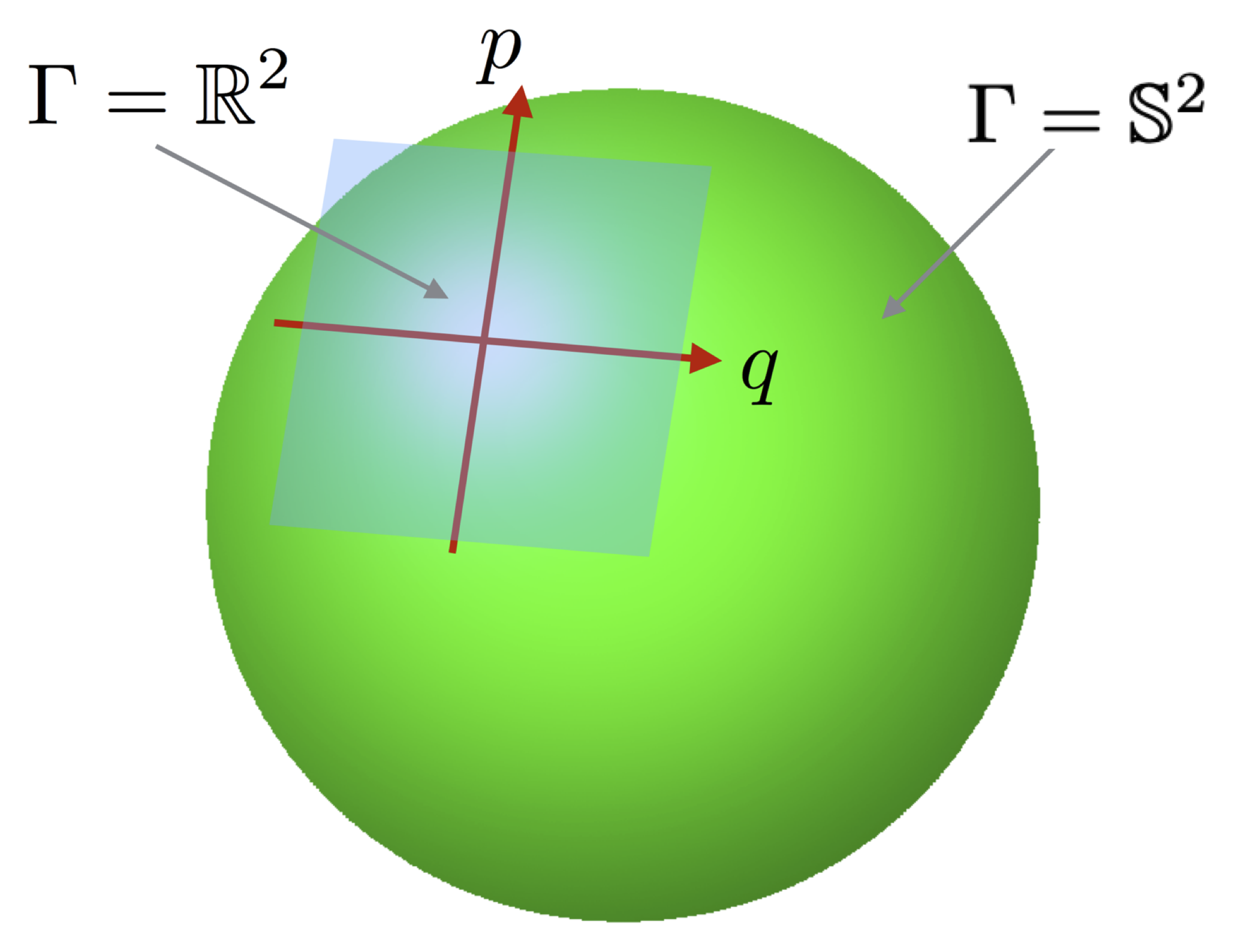}
	\caption{Ilustration of the spherical phase space and its local flat approximation.}
	\label{fig:sphere}
\end{figure}

A natural symplectic form of on the sphere is $\omega_{\mathbb{S}^2}  
= S \sin \theta d \phi \wedge d \theta$, where $\phi \in (-\pi, \pi]$ and $\theta \in [0,\pi]$ 
are spherical angles. It has been shown in \cite{Guimarey:2019lmn} than by 
applying the change of variables  $\varphi = \frac{p}{R_1}$ and 
$\theta = \frac{\pi}{2}+\frac{q}{R_2}$, together with condition $R_1R_2=S$, the 
symplectic form on the 2-sphere takes the form:
\begin{equation}
\omega = \cos\left(\frac{q}{R_2}\right) dp \wedge dq. 
\label{SympSphere}
\end{equation} 
The symplectic form reduces to the flat case in the $S\rightarrow \infty$ ($R_{1,2} \rightarrow \infty$)
limit. Therefore, the symplectic form (\ref{SympSphere}) allows to recover the flat case as a 
local or the large $S$ approximation (see Fig. \ref{fig:sphere}). 

Because the $p$ and $q$ variables do not provide a continuous parametrization 
of the sphere, it is convenient to work withe the spin variables. Here, they correspond
to the following Cartesian parametrization of the 2-sphere:
\begin{align}
S_x&=S\cos\left(\frac{p}{R_1}\right)\cos\left(\frac{q}{R_2}\right), \label{S_x}\\
S_y&=S\sin\left(\frac{p}{R_1}\right)\cos\left(\frac{q}{R_2}\right), \label{S_y}\\
S_z&=-S\sin\left(\frac{q}{R_2}\right),
\label{S_z}
\end{align}
together with the condition $S_x^2+S_y^2+S_z^2=S^2$. Employing the 
symplectic form (\ref{SympSphere}), which defines the Poisson bracket,
one can verify that the spin components satisfy the $\mathfrak{su(2)}$ 
algebra $\{S_i,S_j\}=\epsilon_{ijk}S_k$, where $i,j,k \in \{x,y,z\}$. The 
algebra is quantized in a straightforward manner, leading to the commutator 
algebra $\left[\hat{S}_i,\hat{S}_j\right]= i \hslash \epsilon_{ijk}\hat{S}_k$.
Irreducible representations of the algebra are labeled by the spin $s$,
such that:
\begin{align}
\hat{S}^2 |s,s_z\rangle &= \hslash s(s+1)|s,s_z\rangle, \\
\hat{S}_z |s,s_z\rangle &= \hslash s_z|s,s_z\rangle,
\end{align}
where $s_z = -s,-s+1,.., s_1,s$. The eigenstates states $ |s,s_z\rangle$, 
span the Hilbert space for a given representation, i.e. $\mathcal{H}_s 
= \text{span} \{ |s,-s\rangle, ...,  |s,s\rangle\}$, and $\dim \mathcal{H}_s =2s+1$.
In what follows, for convenience, we set $\hslash=1$.

Let us notice that the parametrization (\ref{S_x}-\ref{S_z}) is not a unique 
choice. In particular, in Ref. \cite{Artigas:2020fab} polar parametrization of a spherical 
phase space was considered. 

Having defined the spherical compactification procedure at the kinematical 
level, let us proceed to dynamics. Te task is now to replace the flat space 
variables in constraints with the spin variables, valid for the spherical phase space. 
It has to be emphasized that the procedure is not unique. However, 
results of various assumptions, satisfying the correspondence to the 
flat phase space, should converge in the large $S$ limit. A simple choice
introduced in \cite{Guimarey:2019lmn} is:  
\begin{align}
p &\rightarrow \frac{S_y}{R_2} = R_1 \sin\left(\frac{p}{R_1}\right)\cos\left(\frac{q}{R_2}\right), \label{defpS} \\
q &\rightarrow -\frac{S_z}{R_1}=  R_2 \sin\left(\frac{q}{R_2}\right),  \label{defqS}
\end{align}
so that the standard case is recovered in the $R_{1,2} \rightarrow \infty$ limit. 
For the system with $N$ degrees of, the same procedure is applied for each 
$(q_i,p_i)$ pair. 

\subsection{de Sitter cosmological model} 

In this article, we will examine application of the compactification procedure 
to the flat de Sitter cosmological model. The gravitational Hamiltonian constraint 
for the model can be written as \cite{Guimarey:2019lmn}:
\begin{equation}
C = q\left(- \frac{3}{4}\kappa p^2+ \frac{\Lambda}{\kappa}\right) \approx 0, 
\label{FRWHamiltonian}
\end{equation}
where $\kappa := 8\pi G $ and $\Lambda$ is a positive cosmological constant. 
Here, the $q$ and $p$ form a canonical pair, for which the symplectic form 
is $\omega = dp \wedge d q$. The $q$ is related with a cubed scale factor, 
so that, the Hubble factor is $H = \frac{1}{3} \frac{\dot{q}}{q}$. By solving the 
constraint, the Friedmann equation is obtained:
\begin{equation}
H^2= \frac{\Lambda}{3}. 
\label{FriedmannFlat}
\end{equation}

In Ref. \cite{Guimarey:2019lmn} a compact phase space generalization of the 
flat de Sitter cosmological model has been introduced. Following the procedure
introduced in the previous subsection, one finds, that the compactified form of 
the constraint (\ref{FRWHamiltonian}) is:
\begin{equation}
C=\frac{S_3}{R_1}\left[\frac{3}{4}\kappa\frac{S_2^2}{R_2^2}-\frac{\Lambda}{\kappa}\right],
\label{HamiltonianSphere}
\end{equation}
where $R_1R_2=S$. By introducing the dimensionless parameter:
\begin{equation}
\delta := \frac{4}{3} \frac{\Lambda}{R_1^2\kappa^2} \in [0,1],
\label{delta}
\end{equation}
and by a proper rescaling, the constraint (\ref{HamiltonianSphere}) can be rewritten 
into the for:
\begin{equation}
C \rightarrow \frac{4S^2}{3 \kappa R_1} C=S_3 S_2^2 -\delta S^2S_3. 
\label{HamiltonianSphere2}
\end{equation}

Quantization of the constrain (\ref{HamiltonianSphere2}), which requires promoting 
of the phase space functions $S_x$, $S_y$ and $S_z$ into operators, and an appropriate 
symmetrization leads to:   
\begin{equation}
\hat{C} = \frac{1}{3}\left( \hat{S}_z\hat{S}_y\hat{S}_y+\hat{S}_y\hat{S}_z\hat{S}_y
+\hat{S}_y\hat{S}_y\hat{S}_z \right)-\delta \hat{S}^2\hat{S}_z. 
\label{QuantumConstraint}
\end{equation}
It has been shown in Ref. \cite{Guimarey:2019lmn} that solutions to the WDW equation 
associated with the constraint (\ref{QuantumConstraint}) can be found. The solutions 
are, however, is not in a direct form but are expressed in terms of the recursion equation. 
Furthermore, solutions to the WDW equation exist for the bosonic representations 
(integer $s$) and, in general, do not exist for the fermionic representations (half-integer $s$).
The first non-trivial solution to the constraint (\ref{QuantumConstraint}) is for $s=1$,
for which the constraint takes the followoing matrix form: 
\begin{equation}
\hat{C}  =2 \left(\frac{1}{6}-\delta\right) \left(\begin{array}{ccc}
1&0&0\\
0&0&0\\
0&0&-1\\
\end{array}\right) =2 \left(\frac{1}{6}-\delta\right) \hat{S}_z. 
\end{equation}
The corresponding solution (excluding the trivial case of $\delta=\frac{1}{6}$) is given 
by the state: 
\begin{align}
|\Psi \rangle = \left(\begin{array}{c} 0\\1\\ 0\\ \end{array}\right) = |s=1, s_z=0\rangle 
                   = \frac{1}{\sqrt{2}}(|01\rangle+|10\rangle), \label{PsiS1}
\end{align} 
where in the last equality, qubit representation of the state is given. 
Therefore, for $s=1$, dimension of the kernel is $\dim \ker \hat{C} =1$.
On the other hand, as discussed in Ref. \cite{Guimarey:2019lmn}, for 
$s=2$ dimension of the kernel depends on the value of $\delta$. 
Namely, for $\delta \neq \frac{7}{18}$ we have $\dim \ker \hat{C} =1$
and for  $\delta = \frac{7}{18}$ we have $\dim \ker \hat{C} =3$. 
Explicit form of the basis states spanning the kernels can be fund in 
Ref. \cite{Guimarey:2019lmn}.

\subsection{Qubit representation}

The introduced compactification, not only leads to finite dimensional Hilbert space, 
but also is suitable for simulations on a quantum computer. Expressing constraint 
with use of spin operators gives a natural qubit representation of the constraint. 
This is because, arbitrary spin $s$ can be decomposed into spin-$1/2$ representations,
which are qubits. One needs $n=2s$ qubits to implement the spin $s$ representation
on a quantum register.  

Moreover, eigenstates of spin operator exhibit symmetry, which can be used to 
simplify ansatz in variational methods. Eigenstates of the $\hat{S}^2$ operator are 
invariant under transformation changing order of qubits, i.e. for the operator 
$\hat{\mathcal{P}}$, defined as:
\begin{equation}
	\hat{\mathcal{P}}|b_1b_2...b_n\rangle=|b_nb_{n-1}...b_1\rangle,
\end{equation}
and if
\begin{equation}
	\hat{S}^2|\psi\rangle=s(s+1)|\psi\rangle,
\end{equation}
we have 
\begin{equation}
	\hat{\mathcal{P}}|\psi\rangle=|\psi\rangle.
\end{equation}

One can simplify the variational ansatz by imposing the $\hat{\mathcal{P}}$ 
symmetry. Using variational methods we need to express our constraint in 
terms of unitary operators. These operators also exhibit the symmetry 
(i.e. $[\hat{C},\hat{\mathcal{P}}]=0$), so we can also reduce number 
of terms for which expectation value must be evaluated. Furthermore, 
utilizing gradient methods to minimize cost function we can use parameter 
shift rule, which can also be optimized according to the symmetry.

\section{Variational solving of a constraint}
\label{VQE}

Following the Dirac quantization method of constrained systems, our task 
it to determine the kernel of the operator $\hat{C}$. The kernel will correspond 
to the physical Hilbert space for the system $\mathcal{H}_{\text{phys}}$ and 
is spanned by the states $|\psi_0\rangle \in \mathcal{H}_{\text{phys}}$, annihilated 
by the Hamiltonian constraint, i.e. satisfying the WDW equation: 
\begin{equation}
	\hat{C}|\psi_0\rangle=0.
\end{equation}

For any linear operator $\hat{C}$, the above condition is equivalent to 
\begin{equation}
\langle\psi_0|\hat{C}^\dagger \hat{C}|\psi_0\rangle=0.
\end{equation}
Moreover, one can proof that 
\begin{equation}
\langle\psi|\hat{C}^\dagger \hat{C}|\psi\rangle\geq 0,
\end{equation}
for all $|\psi\rangle$. In the case of self-adjoint operator $\hat{C}$, the 
corresponding conditions are:  
\begin{align}
&\hat{C}|\psi_0\rangle=0\Longleftrightarrow\langle\psi_0|\hat{C}^2|\psi_0\rangle=0, \\
&\langle\psi|\hat{C}^2|\psi\rangle\geq0.
\end{align}	

Following the Variational Quantum Eigensolver (VQE) methods,   
let us now assume that $|\psi\left(\bm{\alpha}\right)\rangle$ is a state parameterized 
by a vector $\bm{\alpha}=\{\alpha_i\}_{i=1,..,p}$. In order to find $|\psi_0\rangle$ 
we have to find a minimum of non-negative cost function $c\left(\bm{\alpha}\right)$, 
which is defined as follow:
\begin{equation}
	c\left(\bm{\alpha}\right):= \langle\psi\left(\bm{\alpha}\right)|\hat{C}^\dagger \hat{C}|\psi\left(\bm{\alpha}\right)\rangle.
\end{equation}
In case of self-adjoint operator $\hat{C}$, the cost function takes the form:
\begin{equation}
	c\left(\bm{\alpha}\right)=\langle\psi\left(\bm{\alpha}\right)|\hat{C}^2|\psi\left(\bm{\alpha}\right)\rangle.
\end{equation}

\begin{figure}[ht!]
	\centering
	\includegraphics[scale=0.5]{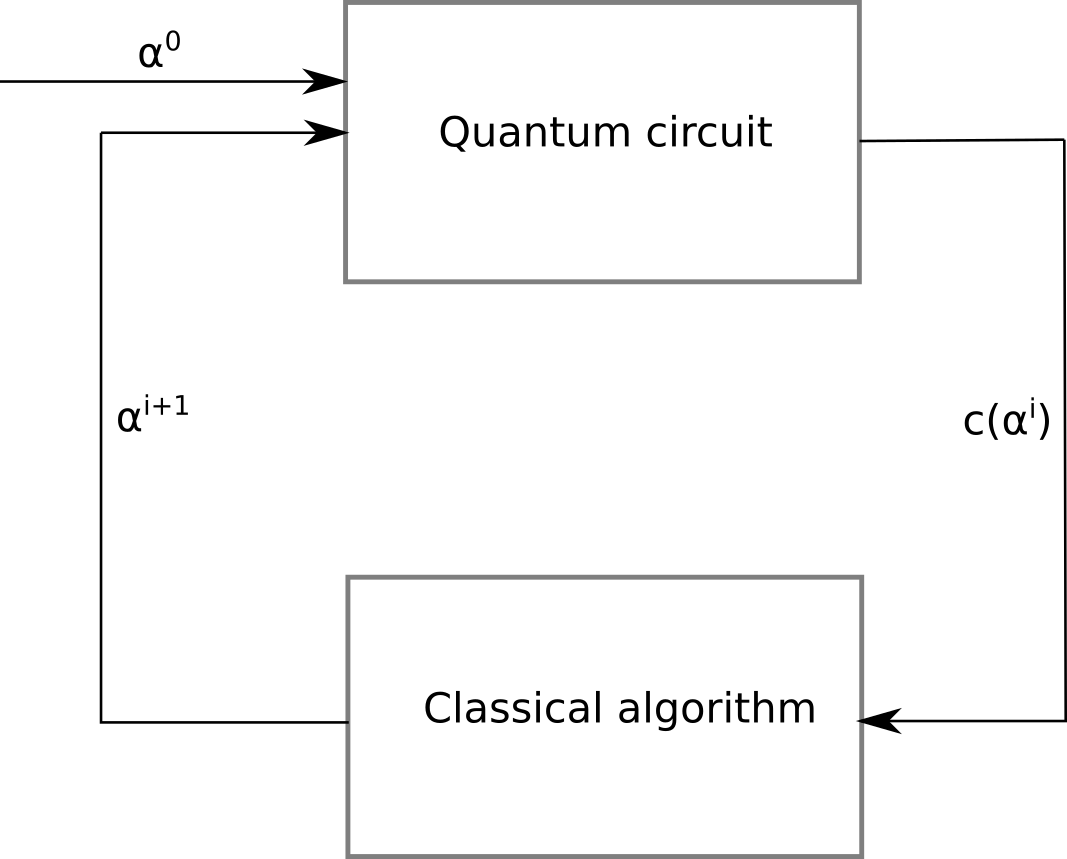}
	\caption{Schematic illustration of the algorithm of finding minimum of cost 
	function, evaluated by a quantum circuit.}
	\label{fig:alg}
\end{figure}

To find minimum of $c$ we use some classical minimizing algorithm 
(on classical a computer) but the value of $c$, i.e. expectation value 
of $\hat{C}^\dagger \hat{C}$ wis computed using a quantum computer.
The algorithm is initialized with some random parameters $\bm{\alpha}_0$. 
Then, by evaluating the quantum circuit, we obtain $c\left(\bm{\alpha_i}\right)$ 
and using classical algorithm we find new parameters $\bm{\alpha_{i+1}}$, 
which are closer to minimum, we repeat this steps (see Fig. \ref{fig:alg}).
Eventually, the set of values:
\begin{equation}
	\bm{\alpha}_{\text{min}} :=\argmin_{\bm{\alpha}}{c(\bm{\alpha})}
	\label{alphamin}
\end{equation}
is found, such that:
\begin{equation}
	|\psi_0\rangle=|\psi\left(\bm{\alpha}_{\text{min}}\right)\rangle.
\end{equation}

Because, in general, the kernel space is more that one dimensional, there 
are different $|\psi_0\rangle$, which satisfy the condition (\ref{alphamin}). 
From the perspective of the operator  $\hat{C}^\dagger \hat{C}$, this reflects 
the fact that its ground sate is degenerated. Therefore, the algorithm must 
be designed such that the whole degeneracy space is sampled.  
  
There are basically two main methods of evaluating the cost function of a 
quantum computer. The first approach utilizes the so-called Hadamard Test 
and the second follows the method discussed in Ref. \cite{Mielczarek:2018jsh}.

\begin{figure}[ht!]
	\leavevmode
	\centering
	\Qcircuit @C=1em @R=1em {
		\lstick{\ket{0}} & \gate{H} & \ctrl{1}         & \qw &\rstick{2\langle\sigma_+\rangle}\\
		\lstick{\ket{0}}& \multigate{2}{V_\psi}      & \multigate{2}{U} & \qw      & \qw  & \qw       \\
		\lstick{\ket{0}}& \ghost{V_\psi}      & \ghost{U}        & \qw      & \qw  & \qw       \\
		\lstick{\ket{0}}& \ghost{V_\psi}      & \ghost{U}        & \qw      & \qw  & \qw       \\
	}
	\caption{The Hadamard test - a circuit measuring expectation value of $\hat{U}$ in state $|\psi\rangle$.}
	\label{fig:exp_val_hadamard_test}
\end{figure}
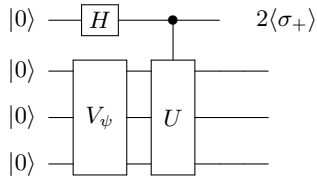

Expectation value of any unitary operator $\hat{U}$ in state $|\psi\rangle$ 
can be measured using Hadamard Test (Fig. \ref{fig:exp_val_hadamard_test}), where
\begin{equation}
\hat{V}_\psi|0\rangle=|\psi\rangle.
\end{equation}
The expectation value of $\hat{U}$ is equal to the expectation value of 
the operator $2\sigma_+=\sigma_x+i\sigma_y$ on the first qubit. When 
$\hat{U}$ is self-adjoint and gives only real expectation values we can 
measure just $\langle\sigma_x\rangle$.

In order to compute $\langle 2\sigma_+\rangle=\langle\sigma_x\rangle+i\langle\sigma_y\rangle$ 
we need to compute $\langle\sigma_x\rangle$ and $\langle\sigma_y\rangle$. First, we need to 
apply gate which allows rotating base from computational one to base of eigenvectors of given 
operator. In the case of $\langle\sigma_x\rangle$, we need to apply Hadamard operator $\hat{H}$ 
and take a measurement of $\sigma_z$ in the compuational bases, then
\begin{equation}
	\langle\sigma_x\rangle=  \langle\hat{H}\sigma_z\hat{H}\rangle = P\left(0\right)-P\left(1\right).
\end{equation}
In the case of $\langle\sigma_y\rangle$, we need to apply operator $\hat{H}\hat{S}^{\dagger}$, where  
$\hat{S}=\left(\begin{array}{cc}
1&0\\
0&i\\
\end{array}\right)$ and take a measurement of $\sigma_z$ in the compuational bases, then 
\begin{equation}
\langle\sigma_y\rangle=\langle\hat{S}\hat{H}\sigma_z\hat{H}\hat{S}^{\dagger}\rangle =  P\left(0\right)-P\left(1\right).
\end{equation}

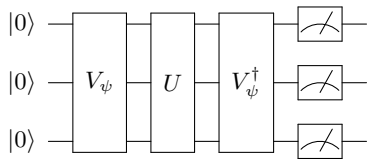
\begin{figure}[ht!]
	\leavevmode
	\centering
	\Qcircuit @C=1em @R=1em {
		\lstick{\ket{0}} & \multigate{2}{V_\psi} & \multigate{2}{U} & \multigate{2}{V_\psi^\dagger} &  \meter & \qw \\
		\lstick{\ket{0}} & \ghost{V_\psi}        & \ghost{U}        & \ghost{V_\psi^\dagger} & \meter & \qw \\
		\lstick{\ket{0}} & \ghost{V_\psi}        & \ghost{U}        & \ghost{V_\psi^\dagger} & \meter & \qw \\
	}
	\caption{Circuit measuring expectation value of $\hat{U}$ in state $|\psi\rangle$.}
	\label{fig:exp_val}
\end{figure}

The another method of evaluating expectation value of a unitary operator without additional qubit
utilizes the the following formula:
\begin{equation}
	\langle\psi| \hat{U} |\psi\rangle=\langle 0|\hat{V}_{\psi}^\dagger \hat{U}\hat{V}_{\psi}|0\rangle.
\end{equation}
So, when we apply the operators $V_\psi$, $U$, $V_\psi^\dagger$ on the initial state $0\rangle$,
and measure probability of a state $|0\rangle$ in the final state we obtain $\left|\langle\psi|\hat{U}|\psi\rangle\right|^2$ 
(see Fig. \ref{fig:exp_val}).

In both methods, the $\hat{U}$ is a unitary operator. In the case of non-unitary operators one has to express 
the operator as a sum of unitary operators and compute expectation value of each unitary term separately.

\section{Computing expectation values}
\label{ExpectationValues}

Let us now proceed to the implementation of the algorithm introduced in the previous 
section. For this purpose, a method of evaluating the mean value of the operator 
$\hat{C}^2$ has to be introduced. 

Our strategy is to express $\hat{C}$ as sum of tensor products of Pauli operators (for the 
spin-1/2 representation):
\begin{equation}
	\hat{C}=\sum_j c_j \bigotimes_i\hat{\sigma}_{ij}^k
\end{equation}
where $k=x,y,z$ indicates one of the Pauli matrices, $i$ is a index of a qubit and $j$ is a index of a term in the sum. 
The $c_j$ are constants multiplying given product of Pauli operators contributing to the sum. 
Then each part of this sum is an unitary operator and expectation value can be easily calculated
\begin{equation}
	\langle \hat{C} \rangle=\sum_j c_j \langle\bigotimes_i \hat{\sigma}_{ij}^k\rangle.
\end{equation}

Here, we apply the method to the constraint (\ref{QuantumConstraint}), which is 
a function of the spin variables $\hat{S}_i$, corresponding to a spin number $s$.
The operators can be expressed in term of the Pauli matrices as followos:
\begin{equation}
	\hat{S}_i=\frac{1}{2}\sum_{j=1}^n\mathbb{I}^1\otimes...\mathbb{I}^{j-1}\otimes\hat{\sigma}_i^j\otimes\mathbb{I}^{j+1}\otimes...\mathbb{I}^n,
\end{equation}
where $n=2s$. It is easy to verify, that the spin components $\hat{S}_i$ obey the 
commutation relation:
\begin{align}
	\left[\hat{S}_i,\hat{S}_j\right]=&\left[\frac{1}{2}\sum_m\hat{\sigma}_i^m,\frac{1}{2}\sum_n\hat{\sigma}_j^n\right]=\frac{1}{4}\sum_{m,n}\left[\hat{\sigma}_i^m,\hat{\sigma}_j^n\right] \nonumber \\
	&=\frac{1}{4}\sum_n\left[\hat{\sigma}_i^n,\hat{\sigma}_j^n\right]=\frac{1}{2}\sum_ni\epsilon_{ijk}\hat{\sigma}_k^n \nonumber \\
	&=i\epsilon_{ijk}\hat{S}_k, \\
\end{align}
where the condition
\begin{equation}
	\left[\hat{\sigma}_i^m,\hat{\sigma}_j^n\right]=0
\end{equation}
for $m\neq n$ has been used. For convenience, we also define:
\begin{widetext}
\begin{align}
&P_n\left(\sigma_i\right) :=\sum_j\mathbb{I}^1\otimes...\mathbb{I}^{j-1}\otimes\sigma_i^j\otimes\mathbb{I}^{j+1}\otimes...\mathbb{I}^n, \\
&P_n\left(\sigma_i,\sigma_j\right) : =\sum_{k,l,k\neq l}\mathbb{I}^1\otimes...\mathbb{I}^{k-1}\otimes\sigma_i^k\otimes\mathbb{I}^{k+1}\otimes...\mathbb{I}^{l-1}\otimes\sigma_j^l\otimes\mathbb{I}^{l+1}\otimes...\mathbb{I}^n, \\
&P_n\left(\sigma_i,\sigma_j,\sigma_p\right) :=\sum_{k,l,q,k\neq l,k\neq q,l\neq q}\mathbb{I}^1\otimes...\sigma_i^k\otimes...\sigma_j^l\otimes...\sigma_p^q\otimes...\mathbb{I}^n,
\end{align}
\end{widetext}
so that we can express
\begin{align}
	\hat{S}_i & =\frac{1}{2}P_n\left(\sigma_i\right), \label{SP} \\
	\hat{S}_i\hat{S}_j &=\frac{1}{4}\left(P_n\left(\sigma_i,\sigma_j\right)+i\epsilon_{ijk}P_n\left(\sigma_k\right)+n\delta_{ij}\mathbb{I}^{\otimes n}\right), \\
	8\hat{S}_i\hat{S}_j\hat{S}_l&=P_n\left(\sigma_i,\sigma_j,\sigma_l\right)+i\epsilon_{ilk}P_n\left(\sigma_k,\sigma_j\right), \nonumber \\
	&+i\epsilon_{jlk}P_n\left(\sigma_k,\sigma_i\right)+i\epsilon_{ijk}P_n\left(\sigma_k,\sigma_l\right) \nonumber \\
	&+\delta_{il}\left(n-1\right)P_n\left(\sigma_j\right)+\delta_{jl}\left(n-1\right)P_n\left(\sigma_i\right) \nonumber \\
	&-\epsilon_{ijk}\epsilon_{klm}P_n\left(\sigma_m\right)+\delta_{ij}nP_n\left(\sigma_l\right)+i\epsilon_{ijl}n\mathbb{I}^{\otimes n}. 
\end{align}

Applying these expressions to Eq. \ref{QuantumConstraint}, we find that: 
\begin{align}
	\hat{C} &= \frac{1}{8}\Bigg(\left(1-\delta\right)P_n\left(\sigma_z,\sigma_y,\sigma_y\right)-\delta P_n\left(\sigma_z,\sigma_x,\sigma_x\right) \nonumber \Bigg.\\
	      &\Bigg. -\delta P_n\left(\sigma_z,\sigma_z,\sigma_z\right)+\left(n-\frac{2}{3}-\delta\left(5n-2\right)\right)P_n\left(\sigma_z\right)\Bigg).\nonumber \\
\end{align}

For the purpose of constructing the cost function, square of the operator $\hat{C}$ has to be evaluated. 
Employing methods of symbolic algebra, the expression for $\hat{C}^2$ can be found explicitly:
\begin{widetext}
\begin{align}
	\hat{C}^2&= \frac{1}{64}\Bigg(P_n\left(\sigma_y,\sigma_y,\sigma_y,\sigma_y,\sigma_z,\sigma_z\right)\left(1-\delta\right)^2
	+P_n\left(\sigma_y,\sigma_y,\sigma_y,\sigma_y\right)\left(n-4\right)\left(1-\delta\right)^2 \nonumber \Bigg.\\
	&+P_n\left(\sigma_y,\sigma_y,\sigma_z,\sigma_z\right)\left(4\left(n-4\right)\left(1-\delta\right)^2-8\delta\left(1-\delta\right)-6\left(n-4\right)\delta\left(1-\delta\right)-12\delta^2+2\left(\frac{3n-2}{3}-\delta\left(5n-2\right)\right)\left(1-\delta\right)\right) \nonumber \\
	&+P_n\left(\sigma_y,\sigma_y\right)\left(4\left(n-2\right)\left(n-3\right)\left(1-\delta\right)^2-8\left(n-2\right)\delta^2+2\left(n-2\right)\left(\frac{3n-2}{3}-\delta\left(5n-2\right)\right)\left(1-\delta\right)\right) \nonumber \\
	&+P_n\left(\sigma_z,\sigma_z\right)\left(2\left(n-2\right)\left(n-3\right)\left(1-\delta\right)^2+20\left(n-2\right)\left(n-3\right)\delta^2+\left(\frac{3n-2}{3}-\delta\left(5n-2\right)\right)^2
	\right. \nonumber \\
	&+ \left. 4\left(n-2\right)\delta\left(1-\delta\right)-6\left(n-2\right)\left(\frac{3n-2}{3}-\delta\left(5n-2\right)\right)\delta\right)  \nonumber \\
	&+\mathbb{I}^{\otimes n}\left(2n\left(n-1\right)\left(n-2\right)\left(1-\delta\right)^2+8n\left(n-1\right)\left(n-2\right)\delta^2+n\left(\frac{3n-2}{3}-\delta\left(5n-2\right)\right)^2\right)
	 \nonumber \\
	&+P_n\left(\sigma_x,\sigma_x\right)\left(4\left(n-2\right)\left(1-\delta\right)^2+4\left(n-2\right)\left(n-3\right)\delta^2+12\left(n-2\right)\delta\left(1-\delta\right)-2\left(n-2\right)\left(\frac{3n-2}{3}-\delta\left(5n-2\right)\right)\delta\right) \nonumber \\
	&+P_n\left(\sigma_x,\sigma_x,\sigma_x,\sigma_x,\sigma_z,\sigma_z\right)\delta^2
	+P_n\left(\sigma_x,\sigma_x,\sigma_x,\sigma_x\right)\left(n-4\right)\delta^2 \nonumber \\
	&+P_n\left(\sigma_x,\sigma_x,\sigma_z,\sigma_z\right)\left(10\left(n-4\right)\delta^2-8\delta\left(1-\delta\right)-2\left(\frac{3n-2}{3}-\delta\left(5n-2\right)\right)\delta+12\delta\left(1-\delta\right)\right)\nonumber \\
	&+P_n\left(\sigma_z,\sigma_z,\sigma_z,\sigma_z,\sigma_z,\sigma_z\right)\delta^2-2P_n\left(\sigma_x,\sigma_x,\sigma_y,\sigma_y,\sigma_z,\sigma_z\right)\delta\left(1-\delta\right)\nonumber\\
	&+P_n\left(\sigma_z,\sigma_z,\sigma_z,\sigma_z\right)\left(9\left(n-4\right)\delta^2-2\left(\frac{3n-2}{3}-\delta\left(5n-2\right)\right)\delta+4\delta\left(1-\delta\right)\right)
	\nonumber \\
	&+P_n\left(\sigma_x,\sigma_x,\sigma_y,\sigma_y\right)\left(-2\delta\left(1-\delta\right)\left(n-4\right)+2\left(1-\delta\right)^2+4\delta^2+8\delta\left(1-\delta\right)\right)
	\nonumber \\
	&\Bigg.-2P_n\left(\sigma_y,\sigma_y,\sigma_z,\sigma_z,\sigma_z,\sigma_z\right)\delta\left(1-\delta\right)+2P_n\left(\sigma_x,\sigma_x,\sigma_z,\sigma_z,\sigma_z,\sigma_z\right)\delta^2\Bigg).
	\label{eq:fullC2}
\end{align}
\end{widetext}

\section{Fixed spin subspace of a quantum register}
\label{FixedSpin}

In order to find the kernel of $\hat{C}$ we have to generate states $|\psi\left(\bm{\alpha}\right)\rangle$. 
In case of 2 qubits we can use circuits allowing to generate any state, but in case of the arbitrary 
many qubits there are no such algorithms and we have to use some ansatz, for example 
$R_y$ or $R_yR_z$ ansatz. Furthermore, often we want to look for the states only in some 
subspace of all possible $n$-qubits states. This is the case in the model under consideration 
for which the constraint (\ref{QuantumConstraint}) is defined in the spin-$s$ subspace of the 
quantum register of $n$-qubits. One possible approach to this issue is generating only 
states obeying a given condition. The possibility is, however, difficult to implement in general.  
Another solution is to generate arbitrary states of the $n$-qubit register and adding a second 
term to the cost function, which fixes a given spin-$s$ subspace.

Let us consider selection of a subspace, which is an eigenspace of some operator $\hat{D}$:
\begin{equation}
	\hat{D}|\psi_d\rangle=d|\psi_d\rangle.
\end{equation}
Then, we have to extend the cost function by adding a term which has minimum value (equal 0) 
for states from this subspace, i.e. 
\begin{align}
	&\langle\psi|\left(\hat{D}-d\mathbb{I}\right)^\dagger\left(\hat{D}-d\mathbb{I}\right)|\psi\rangle \nonumber \\
	&=\langle \hat{D}^\dagger \hat{D}\rangle_{\psi}-2\Re\left(d\langle \hat{D}^\dagger\rangle_{\psi}\right)+\left|d\right|^2.
\end{align}
In consequence, the new cost function takes form:
\begin{align}
	c\left(\bm{\alpha}\right)&=\langle\psi\left(\bm{\alpha}\right)|\hat{C}^\dagger \hat{C}|\psi\left(\bm{\alpha}\right)\rangle \nonumber \\
	&+\langle\psi\left(\bm{\alpha}\right)|\left(\hat{D}-d\mathbb{I}\right)^\dagger\left(\hat{D}-d\mathbb{I}\right)|\psi\left(\bm{\alpha}\right)\rangle.
\end{align}

Due to the numerical minimization issues it is, however, better to consider a normalized cost function. 
In our case, we normalized both terms individually, so the normalized first term is equal:
\begin{equation}
	\frac{1}{\max{\left|\lambda_i\right|^2}}\langle\psi\left(\bm{\alpha}\right)|\hat{C}^\dagger \hat{C}|\psi\left(\bm{\alpha}\right)\rangle
\end{equation}
where $\max{\left|\lambda_i\right|^2}$ is the modulus square of the biggest eigenvalue of $\hat{C}$.
The normalized second term is equal:
\begin{equation}
	\frac{1}{\max{\left|d_i-d\right|^2}}\langle\psi\left(\bm{\alpha}\right)|\left(\hat{D}-d\mathbb{I}\right)^\dagger
	\left(\hat{D}-d\mathbb{I}\right)|\psi\left(\bm{\alpha}\right)\rangle,
\end{equation}
where $d_i$ are eigenvalues of $\hat{D}$. 

In case we do not know these eigenvalues \emph{a priori} we can 
treat them as some parameter that has to be adjusted during simulation 
or we can try to maximize the cost function and in this way estimate 
the value of $\max{\left|\lambda_i\right|^2}$. At the end, we normalize 
both terms dividing them by 2 (we also can take these two terms with 
some different weights than $\frac{1}{2}$ in order to improve performence 
of algorithms in some cases). In this way, the cost function takes values 
from the interval  $[0,1]$:
\begin{equation}
	0\leq c\left(\bm{\alpha}\right)\leq 1.
\end{equation}

In case of the constraint (\ref{QuantumConstraint}), we want to look for 
a kernel in subspace of a given spin $s$.Therefore, we need to take 
cost function with $\hat{D}=\hat{\vec{S}}^2=\hat{D}^\dagger$ and 
$d=s\left(s+1\right)$. However, in this case we can choose simpler 
second term of cost function:
\begin{equation}
	s\left(s+1\right)-\langle\hat{\vec{S}}^2\rangle,
\end{equation}
which after normalization is:
\begin{equation}
	1-\frac{\langle\hat{\vec{S}}^2\rangle}{s\left(s+1\right)}.
\end{equation}

Because (assuming that we use only $n=2s$ qubits) the operator $\vec{S}^2$ 
has maximal expectation value equal to $s\left(s+1\right)$ and minimal equal $0$, so
\begin{equation}
	0\leq 1-\frac{\langle\vec{S}^2\rangle}{s\left(s+1\right)}\leq 1.
\end{equation}
The 0 value corresponds only to the states with spin $s$. So the whole 
cost function has zero value only for states which are simultaneously in 
the kernel of $\hat{C}$ and have spin $s$.  As a spin operator can be 
expressed in terms of qubits, employing Eq. \ref{SP}, we find that: 
\begin{align}
	\vec{S}^2 &=\sum_{i=x,y,z}S_i^2= \frac{3}{4}n\mathbb{I}^{\otimes n}+ \frac{1}{4}P_n\left(\sigma_x,\sigma_x\right) \nonumber \\
	                &+\frac{1}{4}P_n\left(\sigma_y,\sigma_y\right)+\frac{1}{4}P_n\left(\sigma_z,\sigma_z\right),
\end{align}
and consequently:
\begin{align}
	\langle\vec{S}^2\rangle &= \frac{3}{4}n+\frac{1}{4}\langle P_n\left(\sigma_x,\sigma_x\right)\rangle \nonumber \\
	                                     &+\frac{1}{4}\langle P_n\left(\sigma_y,\sigma_y\right)\rangle+\frac{1}{4}\langle P_n\left(\sigma_z,\sigma_z\right)\rangle.
\end{align}

\subsection{Degenerate kernel}

In the case of degenerate kernel (i.e. there are more than one eigenstates for eigenvalue 0) 
we have to first find some eigenstate $ |\psi_1\rangle$ using presented cost function and then 
in order to find another eigenstate (orthogonal to first one) we have to add to the cost function 
following term:
\begin{equation}
	\left|\langle\psi_1|\psi\left(\bm{\alpha}\right)\rangle\right|.
\end{equation}
This term can easily be evaluated using $\langle 0|\hat{V}_{\psi_1}\hat{V}^\dagger_{\psi\left(\bm{\alpha}\right)}|0\rangle$, 
where $\hat{V}_\psi$ generates state $|\psi\rangle$. Using this new cost function we can find 
another state from kernel $|\psi_2\rangle$ and then, adding 
to cost function new term
\begin{equation}
	\left|\langle\psi_2|\psi\left(\bm{\alpha}\right)\rangle\right|,
\end{equation}
we can find the third state, and so on.

Another method is to find different vectors from kernel using algorithm 
starting from different initial parameters and then orthogonalize these 
vectors using the Gramm-Schmidt procedure. When Gramm-Schmidt 
returns zero vector that means that we reached dimension of the kernel 
(or that we generated linearly dependent vector so we have to repeat 
procedure to have high level of confidence that there are no more 
linearly independent vectors in kernel subspace).

\section{Gradient methods}
\label{Gradient}

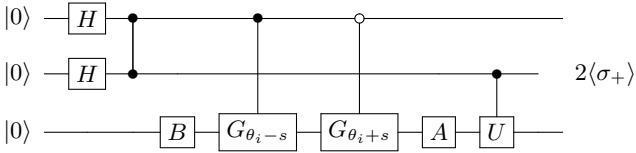
\begin{figure}[ht!]
	\leavevmode
	\centering
	\Qcircuit @C=1em @R=1em {
		\lstick{\ket{0}}&\gate{H}&\ctrl{1}&\qw&\ctrl{2}&\ctrlo{2}&\qw&\qw&\qw&\qw\\
		\lstick{\ket{0}}&\gate{H}&\ctrl{-1}&\qw&\qw&\qw&\qw&\ctrl{1}&\qw&\rstick{2\langle\sigma_+\rangle}\\
		\lstick{\ket{0}}&\qw&\qw&\gate{B}&\gate{G_{\theta_i-s}}&\gate{G_{\theta_i+s}}&\gate{A}&\gate{U}&\qw&\qw\\
	}
	\caption{Parameter-shift rule in the case of one-qubit states and operators.}
	\label{fig:shift_circ}
\end{figure}

In order to find minimum of cost function one needs to use some classical optimizer. 
There are possible choices of optimizers which do not need gradient of cost function, 
for example, COBYLA optimizer \cite{Powell}. But one can gets better performance 
using gradient descent optimizer, for example a basic gradient descent optimizer which 
in each step computes the new values according to the rule:
\begin{equation}
	x^{(t+1)} = x^{(t)} - \eta \nabla f(x^{(t)}).
\end{equation}
Another possibility are some more sophisticated algorithms like the Adam optimizer \cite{Adam}.
Using gradient method one needs to compute gradient of a cost function. One way to do so is
to calculate values of the cost function in two points and compute numerical derivative by 
a finite differences. Another way is to use the parameter-shift rule \cite{Schuld:2019}. Let 
us consider function $f$, which is an expectation value of some operator $\hat{U}$ in some, 
parametrized by $\theta$, state $|\psi\left(\theta\right)\rangle=\hat{V}_\theta|0\rangle$:
\begin{equation}
f\left(\theta\right)=\langle U\rangle_{\psi_\theta}=\langle 0V_\theta^\dagger UV_\theta 0\rangle
\end{equation}
and assume that $V_\theta$ can be factorized (i.e. in circuit representing $V_\theta$ there is one-qubit gate $G_{\theta_i}$):
\begin{equation}
V_\theta=A_{\theta_0,...,\theta_i-1}G_{\theta_i}B_{\theta_i+1,...,\theta_n}
\end{equation}
where
\begin{equation}
G_{\theta_i}=e^{-i\theta_i\mathcal{G}},
\end{equation}
and $\mathcal{G}$ is a self-adjoint operator with two different eigenvalues $-r,+r$. 
Then, exact (not approximate) derivative of $f$ with respect to $\theta_i$ is:
\begin{equation}
\partial_{\theta_i}f=r\left(f\left(\theta_i+s\right)-f\left(\theta_i-s\right)\right)
\end{equation}
where $s=\frac{\pi}{4r}$.
In frequent cases of rotations around Pauli matrices (as in popular ansatzs): 
$G=R_y,R_x,R_z$, parameter $r$ equals $1$. Computation of difference between 
shifted functions can be made in one circuit using ancilla qubit and by controlling 
$G$ gate (see Fig. \ref{fig:shift_circ}). Then, the expectation value of $2\langle\sigma_+\rangle$ 
is equal to derivative of $f$:
\begin{equation}
2\langle\sigma_+\rangle=f\left(\theta_i+s\right)-f\left(\theta_i-s\right)=\partial_{\theta_i}f.
\end{equation}

\section{The $s=1$ case}
\label{S1}

\begin{figure}[ht!]
	\leavevmode
	\centering
	\Qcircuit @C=1em @R=1em {
		\lstick{\ket{0}} & \gate{H} & \ctrl{2} & \ctrl{1} &\qw & \rstick{2\langle\sigma_+\rangle}\\
		\lstick{\ket{0}}& \multigate{1}{V_\psi} & \qw & \gate{Z} & \qw & \qw\\
		\lstick{\ket{0}}& \ghost{V_\psi} & \gate{Z} & \qw & \qw & \qw\\
	}
	\caption{Circuit measuring expectation value $\langle\sigma_z\otimes\sigma_z\rangle$ in state $|\psi\rangle$.}
	\label{fig:szsz}
\end{figure}
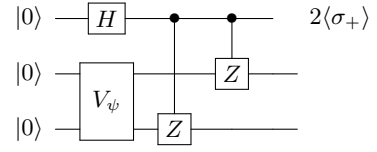

As the first example of the introduced method, let us consider the special case of $s=1$, 
for which the constrain takes form:
\begin{equation}
	\hat{C}=\left(\frac{1}{6}-\delta\right)P_n\left(\sigma_z\right)=
	\left(\frac{1}{6}-\delta\right)\left(\sigma_z\otimes\mathbb{I}+\mathbb{I}\otimes\sigma_z\right),
\end{equation}
and its square
\begin{equation}
	\hat{C}^2=2\left(\frac{1}{6}-\delta\right)^2\left(\mathbb{I}\otimes\mathbb{I}+\sigma_z\otimes\sigma_z\right).
\end{equation}
Therefore, we need to compute only one expectation value:
\begin{equation}
	\langle \hat{C}^2\rangle=2\left(\frac{1}{6}-\delta\right)^2\left(1+\langle\sigma_z\otimes\sigma_z\rangle\right). 
	\label{eq:exp_val_C2_s1}
\end{equation}
The quantum circuit enabling measuring the expectation value 
$\langle\sigma_z\otimes\sigma_z\rangle$ is shown in Fig. \ref{fig:szsz}.

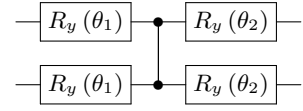
\begin{figure}[ht!]
	\leavevmode
	\centering
	\Qcircuit @C=1em @R=1em {
		& \gate{R_y\left(\theta_1\right)} & \ctrl{1} & \gate{R_y\left(\theta_2\right)}&\qw\\
		& \gate{R_y\left(\theta_1\right)} & \ctrl{-1} & \gate{R_y\left(\theta_2\right)}&\qw\\
	}
	\caption{Quantum circuit for the ansatz for spin 1.}
	\label{fig:s1ansatz}
\end{figure}

As the $\hat{V}_{\psi}$ operator we use the RYCZ ansatz with two angles 
$\theta_1, \theta_2 \in [0, 2\pi)$ (see Appendix A). The quantum circuit for the ansatz 
is shown in Fig. \ref{fig:s1ansatz}. We reduced number of parameters from 
4 to 2 using symmetry of our constraint, which induces symmetry of states 
from the kernel. The states are, namely, invariant under changing qubits order 
from $q_1,q_2,...,q_n$ to $q_n,q_{n-1},...,q_1$.

In this example, basic gradient-descent optimizer was used with stepsize 
$\eta=1$ and convergence tolerance was $10^{-6}$. Parameter $\delta$ in 
constraint was set to $\frac{1}{2}$.

\begin{figure}[ht!]
	\includegraphics[scale=0.35]{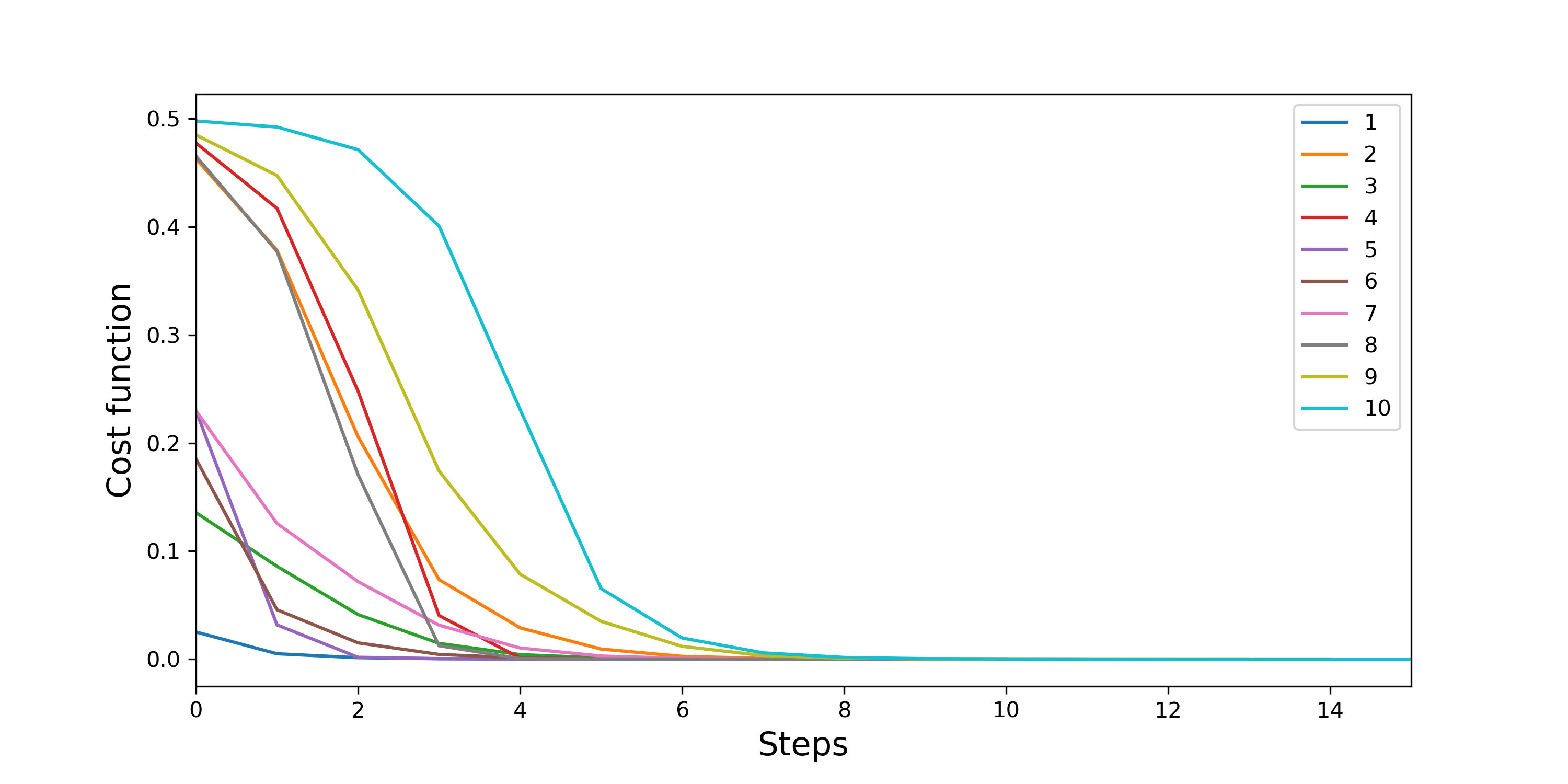}
	\caption{Cost function during minimalization for 10 runs with randomly initialize parameters.}
	\label{fig:cost_spin_1_RY_CX}
\end{figure}

Fig. \ref{fig:cost_spin_1_RY_CX} shows the cost function in a function of 
steps of the optimization procedure.

\begin{figure}[ht!]
	\includegraphics[scale=0.40]{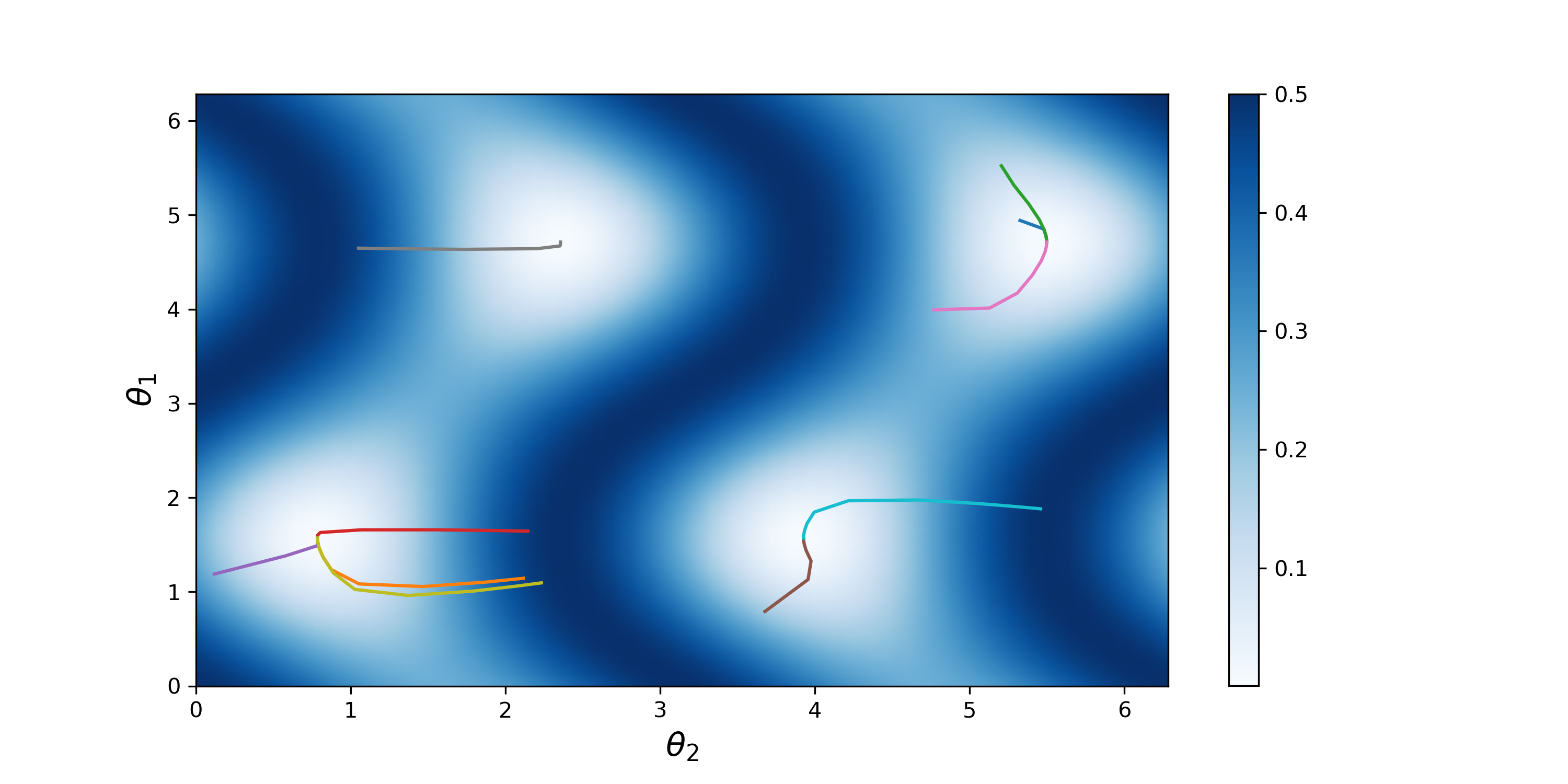}
	\caption{Parameters of states for each step in space of all possible parameters for the case 
	of a quantum simulator. The colors of the curves correspond to those in Fig. \ref{fig:cost_spin_1_RY_CX}. 
	The heatmap represents value of the cost function. }
	\label{fig:cost_heat_map}
\end{figure}

Fig. \ref{fig:cost_heat_map} shows cost function in the parameter space. 
We see four minima, each corresponds to the same state (up to a global 
phase).  It is worth noticing that the landscape of the cost function 
dos not posses any local minima, which simplifies the optimization 
procedure. This, however, not necessarily the case for higher spins, 
including the spin 2 example discussed in the next section. 

\begin{figure}[ht!]
	\includegraphics[scale=0.35]{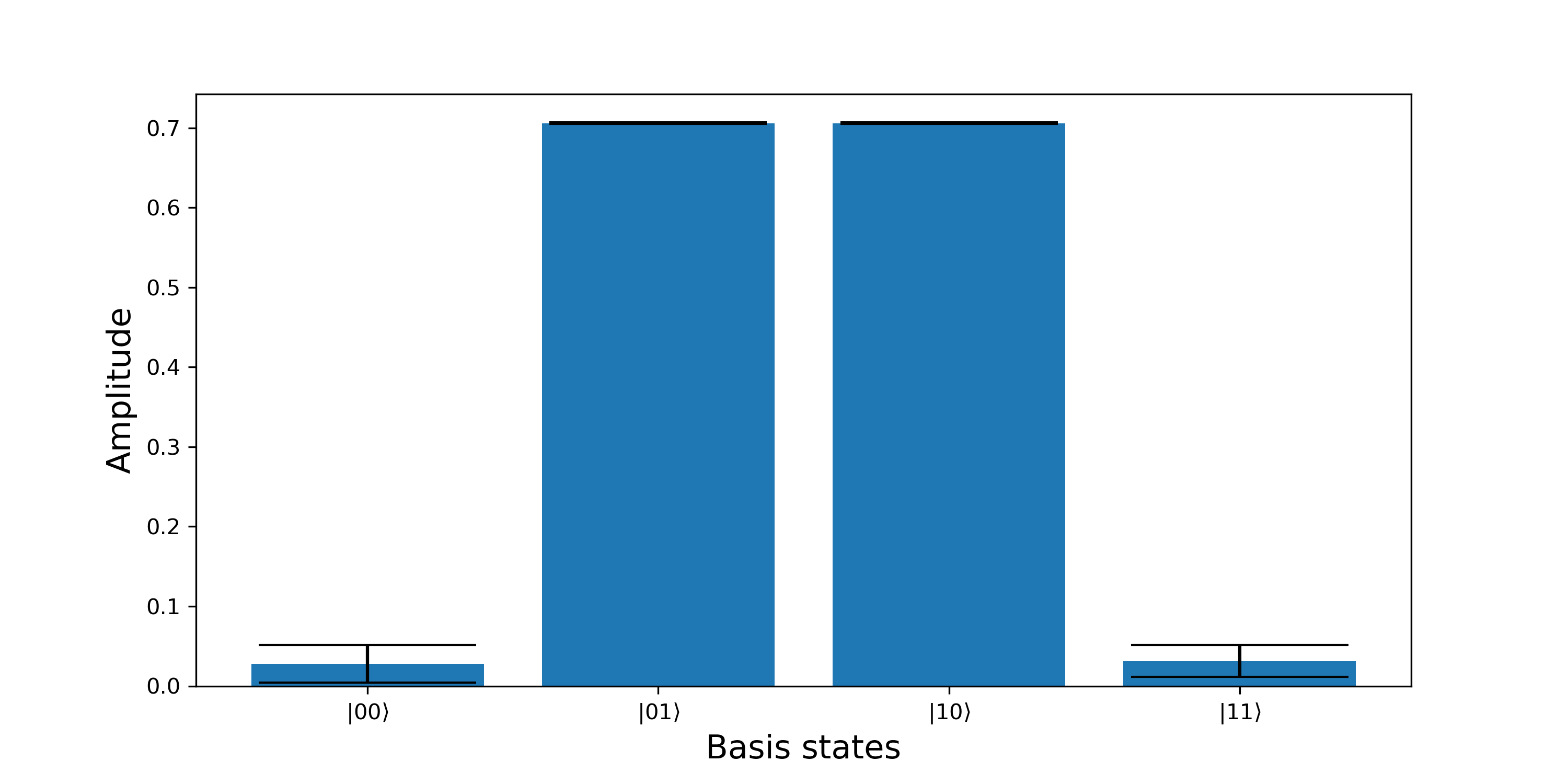}
	\caption{Averaged amplitudes of the final state obtained from over the 10 runs. 
	The RYCZ ansatz implies that the amplitudes are real valued. The error bars 
        correspond to standard deviation.}
	\label{fig:final_states_spin_1_RY_CX}
\end{figure}

Obtained amplitudes of the basis states are shown on Fig. \ref{fig:final_states_spin_1_RY_CX}. 
The algorithm returns correct states up to global phase $\pm1$, on plot signs are agreed. 
The states are very close to exact result, which is a state $\frac{1}{\sqrt{2}}\left(|01\rangle+|10\rangle\right)$
(see Eq. \ref{PsiS1}), with quantum fidelity equal 1, up to numerical uncertainty. Because the RYCZ 
ansatz corresponds to a pure state, the quantum fidelity reduces to $F(\hat{\rho}_1,\hat{\rho}_2) 
= |\langle \psi_1 |\psi_2 \rangle |^2$, where $\hat{\rho}_1 =  |\psi_1 \rangle\langle \psi_1 |$  and
 $\hat{\rho}_2 =  |\psi_2 \rangle\langle \psi_2 |$. 

\begin{figure}[ht!]
	\includegraphics[scale=0.4]{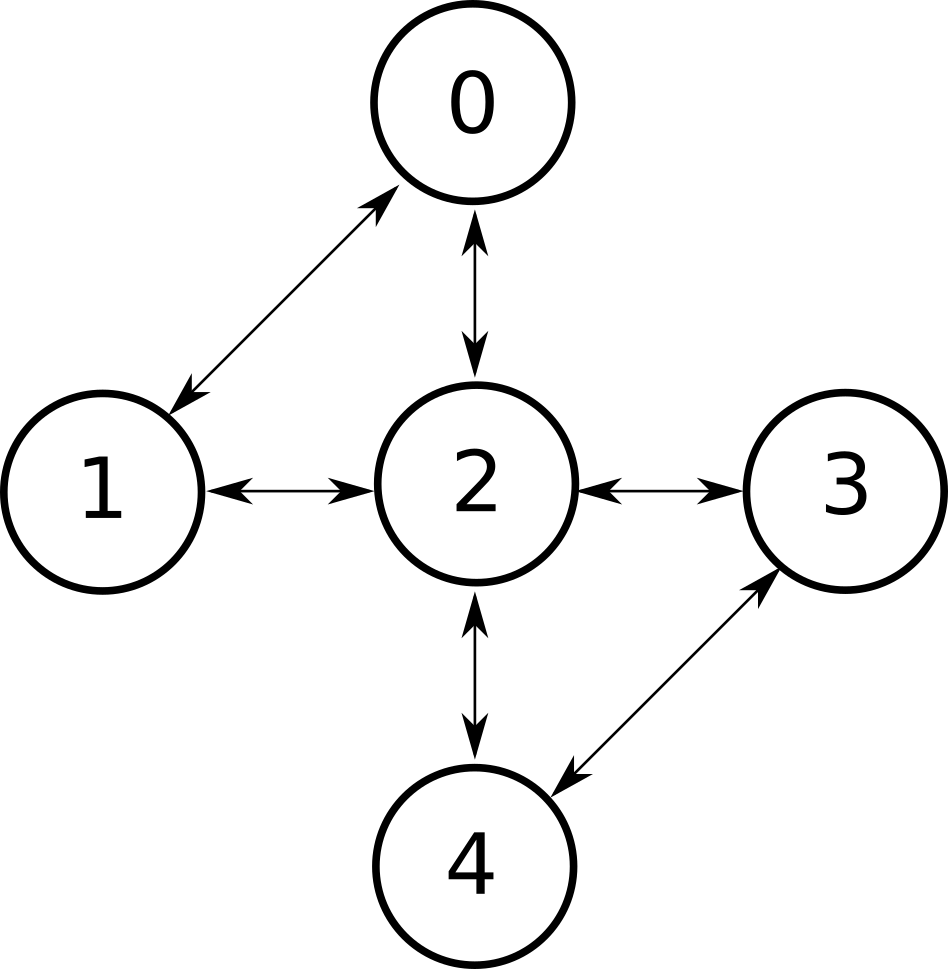}
	\caption{Connectivity of qubits in the Yorktown processor.}
	\label{fig:yorktown_topology}
\end{figure}

For this simple case we also made computation on real IBM superconducting 
quantum processor Yorktown \cite{IBM}. The quantum procesor has topology
of qubits shown on Fig. \ref{fig:yorktown_topology}. In the simulations, 1024 of 
shots for each circuit has been made.  

\begin{figure}[ht!]
	\includegraphics[scale=0.35]{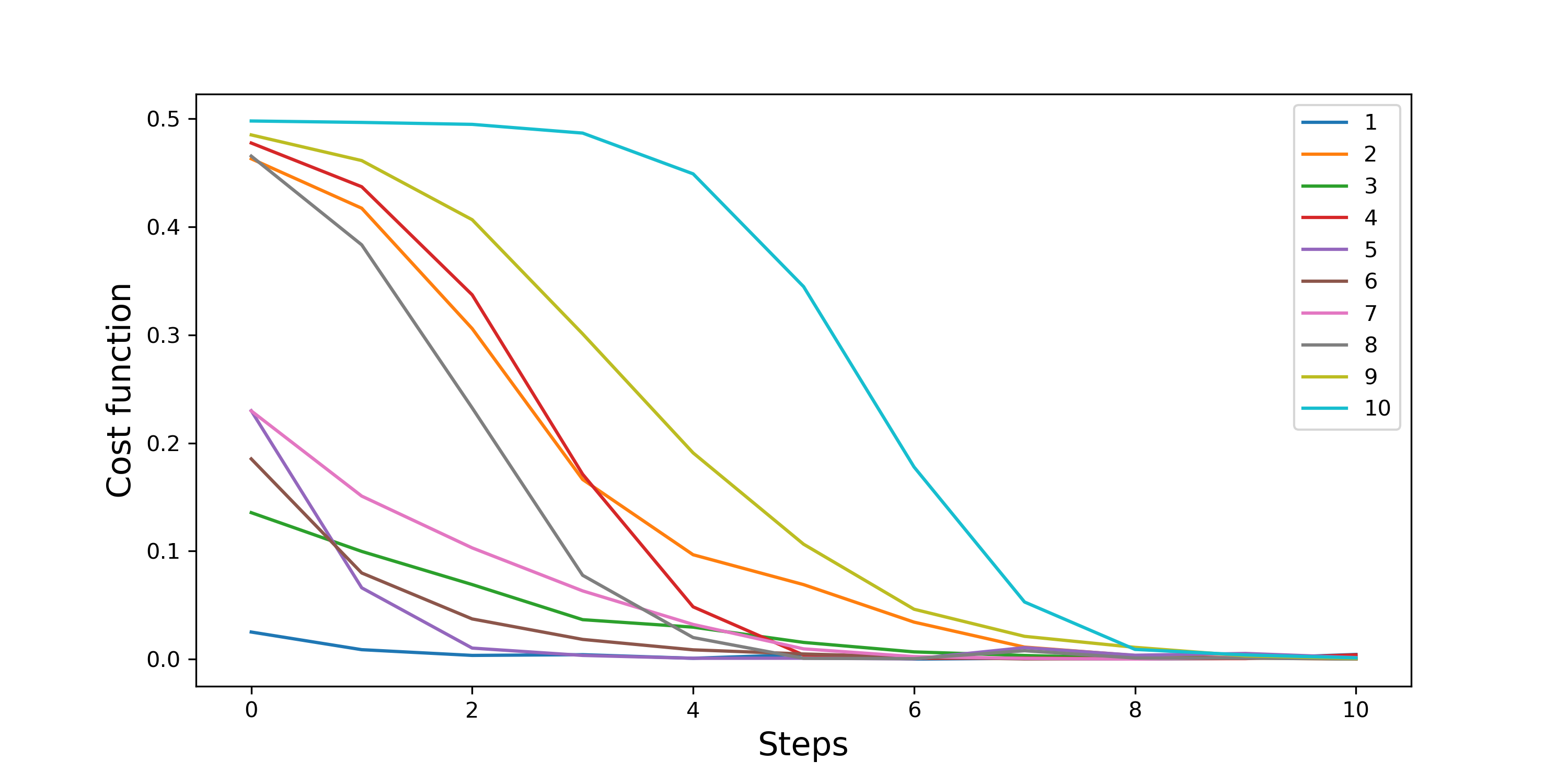}
	\caption{Cost function during minimalization for 10 runs with randomly initialize parameters. 
	Values compute on simulator based on parameters obtained on IBM Yorktown quantum computer.}
	\label{fig:cost_spin_1_RY_CX_qcomp}
\end{figure}

Fig. \ref{fig:cost_spin_1_RY_CX_qcomp} shows the cost function in a function of 
steps of the optimization procedure.

\begin{figure}[ht!]
	\includegraphics[scale=0.40]{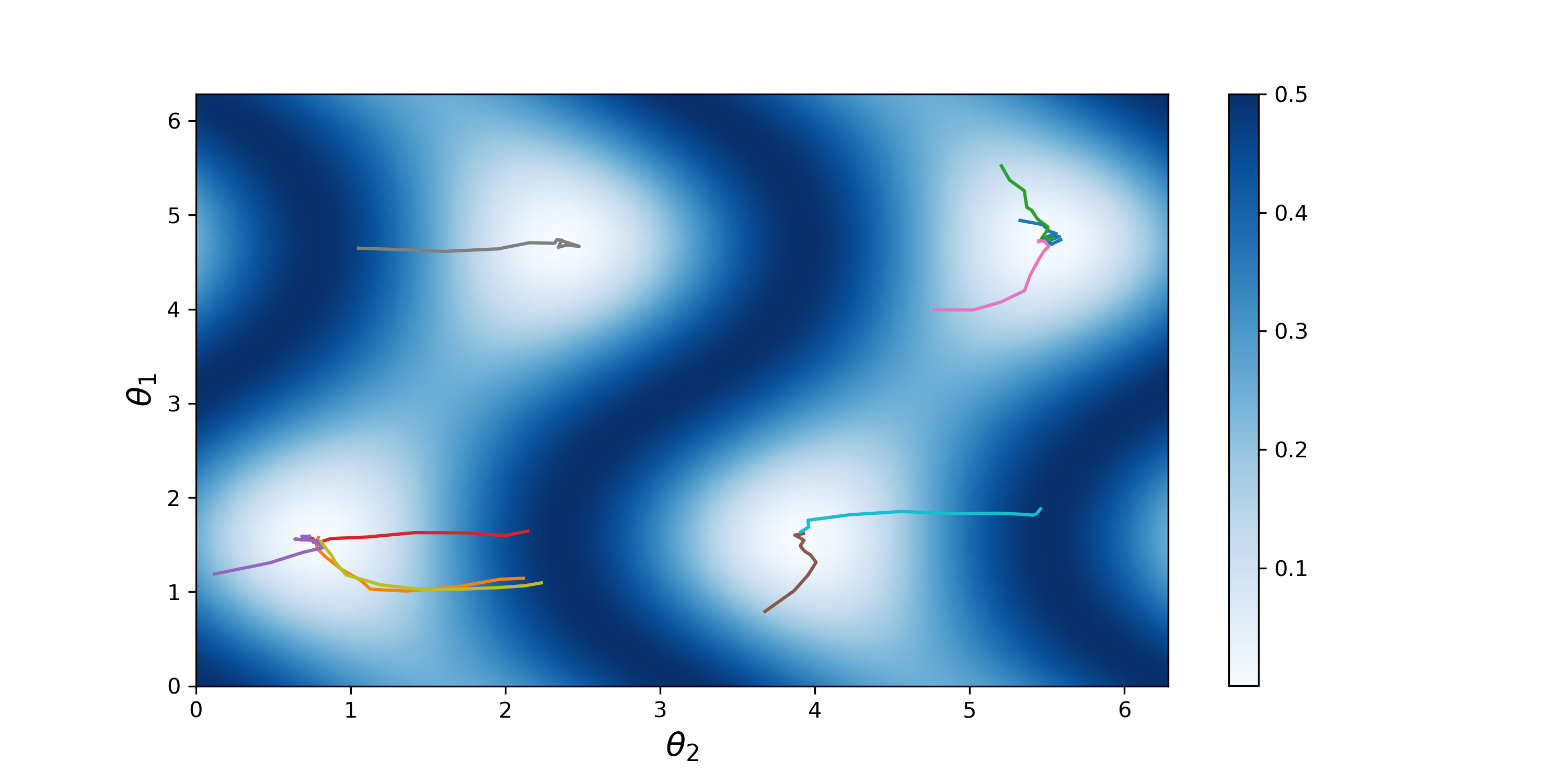}
	\caption{Parameters of states for each step in space of all possible parameters for the case of 
	Yorktown quantum processor. The colors of the curves correspond to those in Fig. 
	\ref{fig:cost_spin_1_RY_CX_qcomp}. The heatmap represents values of the cost function.}
	\label{fig:cost_heat_map_qcomp}
\end{figure}

Fig. \ref{fig:cost_heat_map_qcomp} shows the cost function in the parameter space. 

In consequence of the applied optimization procedure the state, for which 
measured amplitudes are shown in Fig. \ref{fig:final_states_spin_1_RY_CX}, 
has been found. Without error correction methods and with automatic transpilation 
of circuit we obtained fidelity of the found state equal $0.9973\pm 0.0029$.

\section{The $s=2$ case}
\label{S2}

In this case, the constrain squared consists of more Pauli terms, which have to 
be evaluated independently and then summed up:
\begin{align}
	\hat{C}^2=&\frac{1}{16}\Bigg(P_n\left(\sigma_y,\sigma_y\right)\left(\frac{16}{3}-\frac{76}{3}\delta+16\delta^2\right)\Bigg. \nonumber \\
	&+P_n\left(\sigma_z,\sigma_z\right)\left(\frac{34}{9}-40\delta+144\delta^2\right) \nonumber \\
	&+P_n\left(\sigma_x,\sigma_x\right)\left(2-\frac{4}{3}\delta+16\delta^2\right) \nonumber \\
	&+\mathbb{I}^{\otimes 4}\left(\frac{208}{9}-144\delta+384\delta^2\right) \nonumber \\
	&+P_n\left(\sigma_x,\sigma_x,\sigma_y,\sigma_y\right) \nonumber \\
	&+P_n\left(\sigma_z,\sigma_z,\sigma_z,\sigma_z\right)\left(-\frac{2}{3}\delta+8\delta^2\right) \nonumber \\
	&+P_n\left(\sigma_y,\sigma_y,\sigma_z,\sigma_z\right)\left(\frac{5}{3}-\frac{38}{3}\delta+8\delta^2\right) \nonumber \\
	&\Bigg.+P_n\left(\sigma_x,\sigma_x,\sigma_z,\sigma_z\right)\left(-\frac{2}{3}\delta+8\delta^2\right)\Bigg). 
\end{align}

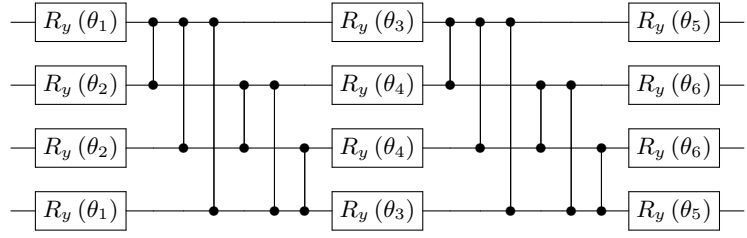
\begin{figure*}[ht!]
	\leavevmode
	\centering
	\Qcircuit @C=1em @R=1em {
		& \gate{R_y\left(\theta_1\right)} & \ctrl{1} & \ctrl{2} & \ctrl{3} & \qw& \qw& \qw& \gate{R_y\left(\theta_3\right)}& \ctrl{1} & \ctrl{2} & \ctrl{3} & \qw& \qw& \qw& \gate{R_y\left(\theta_5\right)} &\qw \\
		& \gate{R_y\left(\theta_2\right)} & \ctrl{-1} & \qw & \qw & \ctrl{1}& \ctrl{2}& \qw& \gate{R_y\left(\theta_4\right)} & \ctrl{-1} & \qw & \qw & \ctrl{1}& \ctrl{2}& \qw&\gate{R_y\left(\theta_6\right)} &\qw\\
		& \gate{R_y\left(\theta_2\right)} & \qw & \ctrl{-2} & \qw& \ctrl{-1} & \qw& \ctrl{1}& \gate{R_y\left(\theta_4\right)} & \qw & \ctrl{-2} & \qw& \ctrl{-1} & \qw& \ctrl{1}&\gate{R_y\left(\theta_6\right)} &\qw\\
		& \gate{R_y\left(\theta_1\right)} & \qw & \qw& \ctrl{-3} & \qw & \ctrl{-2}& \ctrl{-1}& \gate{R_y\left(\theta_3\right)} & \qw & \qw& \ctrl{-3} & \qw & \ctrl{-2}& \ctrl{-1}&\gate{R_y\left(\theta_5\right)} &\qw\\
	}
	\caption{Quantum circuit for the ansatz for spin 2.}
	\label{fig:s2ansatz}
\end{figure*}

Here, we use also the RYCZ ansatz but for 4 qubits, and with 2 layers, and 6 real parameters (see Fig. \ref{fig:s2ansatz}).

\begin{figure}[ht!]
	\includegraphics[scale=0.35]{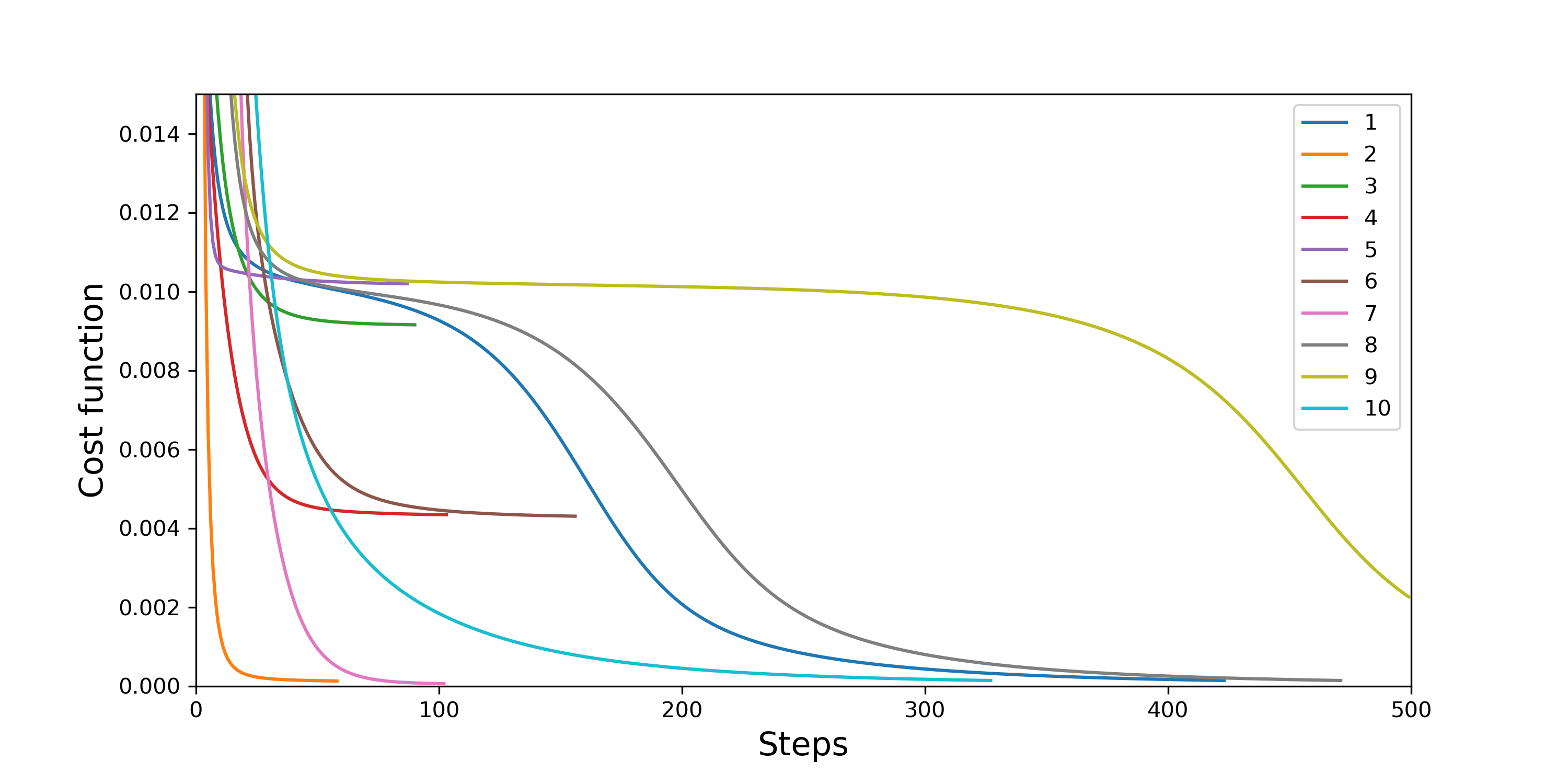}
	\caption{Cost function during minimalization for 10 runs with randomly initialize parameters. 
	Here, the RY-CZ ansatz and $\delta=1/2$ have been used.}
	\label{fig:cost_spin_2_RY_CX_qcomp_b}
\end{figure}

Fig. \ref{fig:cost_spin_2_RY_CX_qcomp_b} shows the cost function in a function of 
steps of the optimization procedure employing simulator of a quantum computer.

In all cases (6 out of 10) when cost function decreases under $0.004$ we 
obtain states with fidelity above $0.98$, in cases (2 out of 10) with cost 
function slightly above $0.04$ we obtain fidelity around $0.64$. Interesting 
is case with high fidelity $0.93$ but with cost function quite high $0.009$, 
which means that even if we stuck in some local minimum we have chance to 
obtain state close to correct one. The obtained fidelities are collected in Tab.
\ref{FidelitiesTab}.

\begin{table}
	\begin{tabular}{|c|c|c|}
		\hline
		No.&Fidelity&Cost\\
		\hline
			1.&0.99825&0.00014\\
			2.&0.99831&0.00013\\
			3.&0.92892&0.00915\\
			4.&0.63477&0.00434\\
			5.&0.00383&0.01019\\
			6.&0.64508&0.00430\\
			7.&0.99777&0.00006\\
			8.&0.99825&0.00014\\
			9.&0.98189&0.00226\\
			10.&0.99825&0.00014\\
		\hline
	\end{tabular}
	\caption{Fidelities of the obtained states and final values of 
	the cost function corresponding to these states for 10 runs.}
	\label{FidelitiesTab}
\end{table}

In the case of $s=2$ for $\delta=\frac{7}{18}$ the kernel of $\hat{C}$ 
is degenerated, $\dim \ker \hat{C} =3$. A linearly independent set 
of states spanning the kernel can be found using the method discussed 
in Sec. \ref{FixedSpin}. In Tab. \ref{Gram-Schmidt} results of the 
Gram-Schmidt procedure for five runs are shown. In all case, there 
are three leading contributions and the fourth one is marginal. 
This provides evidence that the kernel space is three dimensional, 
in accordance to the theoretical expectations for the case with 
$\delta=\frac{7}{18}$. However, due to computational errors, the 
answer is not fully conclusive. 

\begin{table}
	\begin{tabular}{|c|c|c|c|c|}
		\hline
		No.&Step 1.&Step 2.&Step 3.&Step 4.\\
		\hline
		1. & 1.0000 & 0.9998 & 0.0764 & 0.0006\\ 
		2. & 1.0000 & 0.9038 & 0.6065 & 0.0002\\ 
		3. & 1.0000 & 0.1097 & 0.4148 & 0.0040\\ 
		4. & 1.0000 & 0.9984 & 0.0003 & 0.2669\\ 
		5. & 1.0000 & 0.7227 & 0.0000 & 0.8264\\ 
		\hline
	\end{tabular}
	\caption{Norms of states obtained in each step of 
	the Gram-Schmidt procedure (before normalization 
	of the sates) for 5 runs of the optimization procedure
	for the case with $s=2$ and $\delta=\frac{7}{18}$.}
	\label{Gram-Schmidt}
\end{table}

\section{Computational complexity}
\label{Complexity}

As we see the case $s=2$ has many more terms to evaluate than the case $s=1$.  
Consequently, in the large spin limit a problem of enormous number of terms
may arise. To analyze the number of Pauli terms for arbitrary spin $s$, let us 
first notice that the number of different Pauli terms in operator $P_n$ is 
given by the combinatorial factor:
\begin{align}
 \frac{n!}{n_x!n_y!n_z!(n-n_x-n_y-n_z)!},
\end{align}
where $n_k$ is number of Pauli operators $\sigma^k$ and $n$ is number 
of qubits. Using formula (\ref{eq:fullC2}) for $C^2$ for arbitrary spin we obtain 
(for $n\geq 6$) number of Pauli terms, i.e. number of quantum circuits to evaluate is:
\begin{align}
         &\frac{151}{720}n(n-1)(n-2)(n-3)(n-4)(n-5) \nonumber\\
	&+\frac{7}{8}n(n-1)(n-2)(n-3)+\frac{3}{2}n(n-1)
	\sim n^6,
\end{align}
which, fortunately, has polynomial computational complexity 
$\mathcal{O}(n^6)$. Here, symmetries of the constraint (which may slightly 
reduce the number of combinations) were not taken into account.  

Another aspect of computational complexity of the method relates 
to the fact that in the proposed method a subspace of the total 
Hilbert space of the quantum register is used. For a register 
composed of $n$ qubits, the Hilbert space has dimension $2^n$.
However, the constraint (with fixed spin $s$) subspace forms 
a $2s+1=n+1$ dimensional subspace. The remaining $2^n-(n+1)$ 
states of the register are not used. This is unavoidably a waste
of quantum resources, and the week side of the approach. Please 
notice that the spin $s$ scales linearly with the number of qubits. 
So, for instance with 16 qubits we can simulate only a system 
with spin $s=8$.  

The ideal situation would be to utilize the whole Hilbert space of the 
quantum register, so that $2s+1=2^n$. In such case, having 
$n=16$ qubits allows to simulate with $s\sim 3 \cdot 10^4$. Finding 
a method of representing spin operators and constraints in this case 
is an open problem, which will be addressed elsewhere.  

The advantage of the method presented here becomes, however, 
sound while systems with sufficiently high numbers ($m$) of the 
classical degrees of freedom are considered. Then, assuming that 
for every degree of freedom the spin representation is $s$, 
the dimension of the Hilbert space of the composite system is 
$(2s+1)^m$. Every spin consumes $n=2s$ qubits of the quantum 
register, so in total $nm=2sm$ qubits are needed. So, smaller
the $s$ less of the quantum resources are wested, i.e. $2^{2sm}-(2s+1)^m$.
In the limiting case of $s=1/2$, all the quantum resources are utilized
$2^{2\frac{1}{2}m}-\left(2\frac{1}{2}+1\right)^m=0$. In this range,
application of quantum methods prognoses to be advantageous 
over the classical method, since the amount of utilized computational
resources grows exponentially with $m$.

\section{Summary}
\label{Summary}

In this article we have introduced and tested a method of solving of 
Wheeler-DeWitt equation employing variational quantum methods. 
For a single constraint $C$, having $m$ classical degrees of freedom, 
the methods consists of the following main steps:
\begin{enumerate}
\item Replace the kinematical phase space $\Gamma = \mathbb{R}^{2m}$ with $\Gamma = \mathbb{S}^{2m}$.  
\item Express the constraint $C$ in terms of spin variables $C(\vec{S}_1,...,\vec{S}_m)$. This can be 
done by applying the replacement:
\begin{align}
p \rightarrow \frac{S_y}{R_2}, \ \  
q \rightarrow -\frac{S_z}{R_1}, \nonumber
\end{align}
where $R_1R_2=S$, for every canonical pair $(q_i,p_i)$.
\item Perform canonical quantization and symmetrization of the constraint, obtaining  
$\hat{C}(\hat{\vec{S}}_1,...,\hat{\vec{S}}_m)$. Fix a particular representation $s$ for the spins. 
\item Represent the spin operators in terms of qubits, employing the formula: 
\begin{equation}
	\hat{S}_i=\frac{1}{2}\sum_{j=1}^n\mathbb{I}^1\otimes...\mathbb{I}^{j-1}\otimes\hat{\sigma}_i^j\otimes\mathbb{I}^{j+1}\otimes...\mathbb{I}^n, \nonumber
\end{equation}
where $n=2s$.
\item Apply the the VQE method with the cost function:
\begin{align}
	c\left(\bm{\alpha}\right) &= \frac{a}{\max{\left|\lambda_i\right|^2}}\langle\psi\left(\bm{\alpha}\right)|\hat{C}^\dagger \hat{C}|\psi\left(\bm{\alpha}\right)\rangle \nonumber \\
	&+ b\left(1-\frac{\langle\hat{\vec{S}}^2\rangle}{s\left(s+1\right)}\right), \nonumber
\end{align}
where $a+b=1$, and $a,b \in (0,1)$ (e.g. $a=\frac{1}{2}=b$).
\item Explore degeneracy of the kernel space by either adding terms 
$\left|\langle\psi_i|\psi\left(\bm{\alpha}\right)\rangle\right|$
to the cost function or by applying the Gramm-Schmidt procedure.
\item Study the large $s$ limit to recover results for the case of the 
flat (affine) phase space. 
\end{enumerate}

The procedure utilizes compactification of the system's phase space 
for the purpose of making its Hilbert space finite. The dimension of 
the Hilbert space is controlled by a single parameter $s$, which labels
irreducible representations of the $SU(2)$ group. The flat phase space
case is recovered in the large spin $s$ limit. In the article, the procedure 
has been tested on the example of a de Sitter cosmological model, 
which has a single classical degree of freedom (the scale factor) and 
consequently two-dimensional phase space. In the quantum case,
the the kinematics is described by the spin operators $\hat{\vec{S}}$.
Quantum constraint of the system has been expressed in terms 
of the action qubits of the quantum register for an arbitrary spin $s$. 
This allowed to perform the VQE method and extract physical states.  

As an example, the procedure has been executed for $s=1$ and 
$s=2$. In the case of $s=1$ the quantum circuits were evaluated 
on both simulator of the quantum computer and superconducting 
quantum computer Yorktown. In case $s=2$ the computations have
been performed on a simulator only due to high quantum errors. 
Both the case of non-degenerate and degenerate kernel was 
explored, confirming correctness of the method. 

As it has been emphasized, the method introduced do not provide 
advantage over classical computations for the case of a single 
degree of freedom (relevant in homogeneous and isotropic cosmology). 
However, the advantage is expected while a large number of 
quantum-gravitational degrees of freedom is considered. Investigation 
of such a case will be a subject of our further studies.  

\section*{Acknowledgments}

The research has been supported by the Sonata Bis Grant No. 
DEC-2017/26/E/ST2/00763 of the National Science Centre Poland. 
This research was funded by the Priority Research Area Digiworld 
under the program Excellence Initiative -- Research University at 
the Jagiellonian University in Krak\'ow. Furthermore, this publication 
was made possible through the support of the ID\# 61466 grant 
from the John Templeton Foundation, as part of the ``The Quantum 
Information Structure of Spacetime (QISS)'' Project (qiss.fr). The 
opinions expressed in this publication are those of the authors 
and do not necessarily reflect the views of the John Templeton 
Foundation.

\section*{Appendix A}

RYCZ ansatz consists of $R_y$ (RY):
\begin{equation}
	R_y\left(\theta\right)=\exp\left(-i\frac{\theta}{2}\sigma_y\right)=\left(\begin{array}{cc}
	\cos\frac{\theta}{2} & -\sin\frac{\theta}{2}\\
	\sin\frac{\theta}{2} & \cos\frac{\theta}{2}\\
	\end{array}\right),
\end{equation}
and controlled-$\sigma_z$ (CZ) gates.

\begin{figure}[ht!]
	\leavevmode
	\centering
	\Qcircuit @C=1em @R=1em {
		& \gate{R_y\left(\theta_1\right)} & \ctrl{1} & \ctrl{2} & \qw & \gate{R_y\left(\theta_4\right)} & \qw \\
		& \gate{R_y\left(\theta_2\right)} & \ctrl{-1} & \qw & \ctrl{1} & \gate{R_y\left(\theta_5\right)} &\qw\\
		& \gate{R_y\left(\theta_3\right)} & \qw & \ctrl{-2} & \ctrl{-1} & \gate{R_y\left(\theta_6\right)} &\qw\\
	}
	\caption{Quantum circuit for the  RY ansatz with full entanglement, and depth $=1$.}
	\label{fig:RYall}
\end{figure}
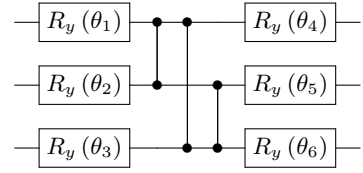

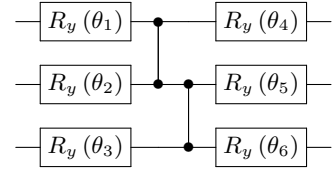
\begin{figure}[ht!]
	\leavevmode
	\centering
	\Qcircuit @C=1em @R=1em {
		& \gate{R_y\left(\theta_1\right)} & \ctrl{1}  & \qw & \gate{R_y\left(\theta_4\right)} &\qw\\
		& \gate{R_y\left(\theta_2\right)} & \ctrl{-1} & \ctrl{1} & \gate{R_y\left(\theta_5\right)} &\qw\\
		& \gate{R_y\left(\theta_3\right)} & \qw & \ctrl{-1} & \gate{R_y\left(\theta_6\right)} &\qw\\
	}
	\caption{Quantum circuit for the RY ansatz with linear entanglement, and depth $=1$.}
	\label{fig:RYlinear}
\end{figure}

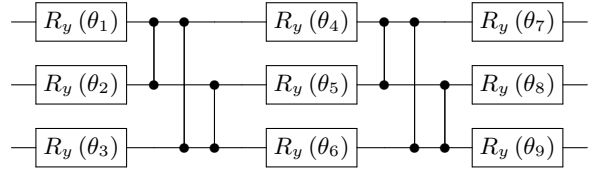
\begin{figure}[ht!]
	\leavevmode
	\centering
	\Qcircuit @C=1em @R=1em {
		& \gate{R_y\left(\theta_1\right)} & \ctrl{1} & \ctrl{2} & \qw & \qw& \gate{R_y\left(\theta_4\right)} & \ctrl{1} & \ctrl{2} & \qw & \gate{R_y\left(\theta_7\right)} &\qw \\
		& \gate{R_y\left(\theta_2\right)} & \ctrl{-1} & \qw & \ctrl{1} & \qw& \gate{R_y\left(\theta_5\right)} & \ctrl{-1} & \qw & \ctrl{1} & \gate{R_y\left(\theta_8\right)} &\qw\\
		& \gate{R_y\left(\theta_3\right)} & \qw & \ctrl{-2} & \ctrl{-1} & \qw& \gate{R_y\left(\theta_6\right)} & \qw & \ctrl{-2} & \ctrl{-1} & \gate{R_y\left(\theta_9\right)} &\qw\\
	}
	\caption{Quantum circuit for the RY ansatz with full entanglement, and depth $=2$.}
	\label{fig:RYall2}
\end{figure}

We apply gates $R_y\left(\theta_i\right)$, parameterized by different parameters $\theta_i$, 
on every qubit and then we apply CZ gates on all pairs of qubits (Fig. \ref{fig:RYall}) or 
only on some pairs (Fig. \ref{fig:RYlinear}). This block of gates can be repeated many 
times (Fig. \ref{fig:RYall2}).

This ansatz generates states with real coefficients, but since $\hat{C}$ is self-adjoint, 
we can always choose eigenvectors to be real. Let $|v\rangle$ be eigenvector with 
complex coefficients for $\lambda$ eigenvalue (which is real):
\begin{equation}
	\hat{C}|v\rangle=\lambda |v\rangle.
\end{equation}
Then $|\bar{v}\rangle = |v\rangle^*$ is also eigenvector for the same eigenvalue:
\begin{equation}
	\hat{C}|\bar{v}\rangle=\lambda |\bar{v}\rangle.
\end{equation}
In consequence, we can take linear combination of these vectors:
\begin{align}
	|a\rangle &=\frac{1}{2}\left(|v\rangle+|\bar{v}\rangle\right), \\
	|b\rangle &=\frac{1}{2i}\left(|v\rangle-|\bar{v}\rangle\right),
\end{align}
which are real eigenvectors to the eigenvalue $\lambda$.

Ansatz from Sec. \ref{S1} can be expressed as RYCZ ansatz by:
\begin{figure}[ht!]
	\centering
	\begin{subfigure}{0.4\textwidth}
		\leavevmode
		\Qcircuit @C=0.7em @R=1em {
			& \gate{R_y\left(\theta_1\right)} & \ctrl{1}  & \qw & \gate{R_y\left(\theta_3\right)} &\qw\\
			& \gate{R_y\left(\theta_2\right)} & \ctrl{-1} & \qw & \gate{R_y\left(\theta_4\right)} &\qw\\
		}
	\end{subfigure}  \\ \vspace{10pt}
	$\equiv$  \vspace{10pt} \\
	\centering
	\begin{subfigure}{0.4\textwidth}
		\leavevmode
		\Qcircuit @C=0.7em @R=1em {
			& \gate{R_y\left(\theta_1\right)} & \ctrl{1}  & \qw & \gate{R_y\left(\theta_3\right)} &\qw\\
			& \gate{R_y\left(\theta_2+\frac{\pi}{2}\right)} & \targ & \qw & \gate{R_y\left(\theta_4-\frac{\pi}{2}\right)} &\qw\\
		}
	\end{subfigure}
\caption{Quantum circuit for the RYCZ ansatz.}
\label{fig:RYCZandSpin1}
\end{figure}
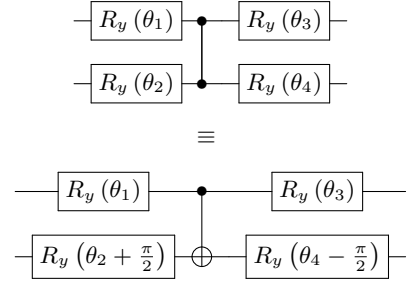

\end{document}